\documentclass[]{emulateapj}

\usepackage{graphicx}
\usepackage{amsmath,amssymb}
\usepackage{natbib}

\newcommand{\HI}{H\,{\sc i}}

\newcommand{\Ha}{H$\alpha$}

\newcommand{\Lyb}{Ly$\beta$}

\newcommand{\kappaB}{\kappa_\mathrm{B}}
\newcommand{\vs}{v_{s}}
\newcommand{\va}{v_{\mathrm{A}}}
\newcommand{\aG}{a_{G}}

\newcommand{\aCG}{a_{C+G}}
\newcommand{\tacc}{t_{\mathrm{SNR}}}

\newcommand{\gamg}{\gamma_{G}}
\newcommand{\gamc}{\gamma_{C}}
\newcommand{\PG}{P_{G}}
\newcommand{\PC}{P_{C}}

\newcommand{\SG}{S_{G}}
\newcommand{\SC}{S_{C}}
\newcommand{\SCa}{S_{C\mathrm{a}}}
\newcommand{\SCi}{S_{C\mathrm{i}}}

\newcommand{\kms}{\,\mathrm{km}\,\mathrm{s}^{-1}}
\newcommand{\cms}{\,\mathrm{cm}^{2}\,\mathrm{s}^{-1}}
\newcommand{\cm}{\,\mathrm{cm}}

\newcommand{\yr}{\,\mathrm{y}}
\newcommand{\Kv}{\,\mathrm{K}}

\newcommand{\keV}{\,\mathrm{keV}}

\newcommand{\kpc}{\,\mathrm{kpc}}

\newcommand{\cmq}{\,\mathrm{cm}^{-3}}

\newcommand{\muG}{\,\mu\mathrm{G}}
\newcommand{\ahour}{^\mathrm{h}}
\newcommand{\aminute}{^\mathrm{m}}
\newcommand{\asecond}{^\mathrm{s}}

\shorttitle{Cosmic-ray precursor of a Balmer-dominated shock}
\shortauthors{Wagner et~al.}

\begin{document}

\title{A cosmic-ray precursor model for a Balmer-dominated shock in Tycho's supernova remnant}

\author{A.~Y.~Wagner\altaffilmark{1}\altaffilmark{2}, J.-J.~Lee\altaffilmark{3}, J.~C.~Raymond\altaffilmark{4}, T.~W.~Hartquist\altaffilmark{1}, and S.~A.~E.~G.~Falle\altaffilmark{1}}

\altaffiltext{1}{School of Physics and Astronomy, University of Leeds, Leeds LS3 1NS, UK; ayw@ast.leeds.ac.uk.}
\altaffiltext{2}{Current address: Research School of Astronomy and Astrophysics, Mount Stromlo Observatory, Australian National University, Cotter Road, Weston Creek, ACT 2611, Australia.}
\altaffiltext{3}{Department of Astronomy and Astrophysics, Pennsylvania State University, 525 Davey Laboratory, University Park, PA 16802.}
\altaffiltext{4}{Harvard-Smithsonian Center for Astrophysics, 60 Garden Street, Cambridge, MA 02138.}

\begin{abstract}
We present a time-dependent cosmic-ray modified shock model for which the calculated \Ha{} emissivity profile agrees well with the \Ha{} flux increase ahead of the Balmer-dominated shock at knot g in \object[Tycho SNR]{Tycho's supernova remnant}, observed by \citet{Lee-etal2007}. The backreaction of the cosmic ray component on the thermal component is treated in the two-fluid approximation, and we include thermal particle injection and energy transfer due to the acoustic instability in the precursor. The transient state of our model that describes the current state of the shock at knot g, occurs during the evolution from a thermal gas dominated shock to a smooth cosmic-ray dominated shock. Assuming a distance of $2.3\kpc$ to Tycho's remnant we obtain values for the cosmic ray diffusion coefficient, $\kappa$, the injection parameter, $\epsilon$, and the time scale for the energy transfer, $\tau$, of $\kappa=2\times10^{24}\cms$, $\epsilon=4.2\times10^{-3}$, and $\tau=426\yr$, respectively. We have also studied the parameter space for fast ($300\kms\lesssim\vs\lesssim3000\kms)$, time-asymptotically steady shocks and have identified a branch of solutions, for which the temperature in the cosmic ray precursor typically reaches $2$--$6\times10^{4}\Kv$ and the bulk acceleration of the flow through the precursor is less than $10\kms$. These solutions fall into the low cosmic ray acceleration efficiency regime and are relatively insensitive to shock parameters. This low cosmic ray acceleration efficiency branch of solutions may provide a natural explanation for the line broadening of the \Ha{} narrow component observed in non-radiative shocks in many supernova remnants.
\end{abstract}


\section{Introduction}
Balmer-dominated filaments in supernova remnants (SNRs) trace fast, non-radiative shocks that propagate into partially neutral, diffuse media. The filaments are sheets of shocked gas seen edge-on \citep{Hester1987} and have been observed and studied in many SNRs including \object[Tycho SNR]{Tycho's SNR} (\citealp{CKR1980,KWC1987,SKBW1991}; \citealp[][hereafter G00 and G01, respectively]{GRHB2000,GRSH2001}; \citealp{Lee-etal2007}, hereafter L07), the Cygnus Loop \citep[e.g.][]{RBFG1983,HRB1994}, RCW86 \citep[e.g.][]{SGLS2003}, Kepler's SNR \citep[e.g.][]{BLV1991,Sankrit-etal2005}, SN1006 \citep[e.g.][]{GWRL2002}, and four remnants in the LMC \citep[e.g.][]{SRL1994}. This work focuses on modeling a particular Balmer-dominated shock located at knot g in Tycho's SNR \citep[after][]{KvB1978} and recently observed by L07. 

The \Ha{} spectral line profile of a Balmer-dominated filament has narrow and broad components, whose widths represent the preshock temperature of neutral H and the postshock temperature of protons, respectively \citep{CR1978}. The width of the broad component and the ratio of intensities are diagnostics for the shock speed and the degree of electron-ion temperature equilibration behind the shock. Shock models to quantify these diagnostics were first developed by \citet{CKR1980} and later improved by \citet{SKBW1991}, G01, \citet{HMcC2007}, and \citet{HAMR2007}. Currently, the most advanced non-radiative shock models are those of \citet{AHMR2008}. Except for the work by \citet{BC1988}, shock models used to interpret optical observations, to date, have not included modifications of the shock structure due to diffusive shock acceleration (DSA) of cosmic rays (CRs). 

Balmer-dominated filaments are usually associated with forward shocks of the expanding SNR bubbles. The only remnant in which Balmer emission from a reverse shock has been observed is SN 1987 \citep{Michael-etal2003}. Forward shocks in SNRs are also sites of CR electron acceleration, as evidenced by synchrotron radio and X-ray emission \citep{Koyama-etal1995,Gotthelf-etal2001,Long-etal2003,Bamba-etal2005,CHBD2007}, as well as likely sites of CR ion acceleration \citep{BE1987,Drury-etal2001SSRv, Warren-etal2005}. In most cases, Balmer-dominated shocks do not show evidence for synchrotron emission, which suggests that particle acceleration in Balmer-dominated shocks is not efficient. The neutral component in the upstream thermal gas that is required for non-radiative shocks to produce Balmer-dominated filaments may be damping the turbulence necessary for efficient cosmic ray acceleration in SNR shocks \citep{DDK1996}. Conversely, the heating of the upstream medium due to efficient CR acceleration may prevent most neutrals reaching the shock before being ionized \citep{HRB1994}. Two known exceptions where Balmer emission and synchrotron X-ray emission coincide are knot g in Tycho's remnant and a small portion of the eastern rim of SN 1006 \citep{Cassam-Chenai-etal2008}. However, a direct connection between the Balmer emission producing shock and the X-ray producing shock cannot be made for either case because the X-ray morphology is not resolved to the level of the optical emission and because the effects of projection are uncertain. Many Balmer-dominated filaments are observed to bound regions of X-ray emission whose spectra are consistent with thermal emission of a shock-heated ambient medium \citep{HRB1994,Raymond-etal2007}.

The theory of DSA predicts a CR precursor ahead of the gas subshock \citep{D-V1981,BE1999}, in which the upstream gas is pre-heated and accelerated over a characteristic distance $\kappa/\vs$, where $\kappa$ is a momentum averaged CR diffusion coefficient, and $\vs$ is the shock speed. The value of $\kappa$ for CR ions depends on the spectrum of the magnetic wave-field, thought to be generated by the CR streaming instability \citep{Bell1978a}, and has not been well constrained by observations yet. While in the general ISM $\kappa\sim3$--$5\times10^{28}\cms$ \citep{SMP2007CR}, at the forward shocks of SNRs $\kappa$ is thought to be close to the Bohm diffusion limit $\kappaB\approx3\times10^{22}\,\mathrm{cm}^2\,\mathrm{s}^{-1}\,\left(\beta\,B_0/\mu{}\mathrm{G}\right)^{-1}\left(p/\mathrm{GeV}\,\mathrm{c}^{-1}\right)$, where $\beta$ is the ratio of particle speed to the speed of light, $B_0$ is the large scale magnetic field, and $p$ is the CR particle momentum \citep{Drury1983}. $\kappaB$ is the lower limit to $\kappa$ allowed by the standard theory of DSA, and corresponds to saturated field fluctuations, $\delta{}B/B_0=1$. \citet{SGLS2003} estimated upper limits for the diffusion coefficient of CR ions in several SNRs in the range $\kappa\sim10^{25}\cms$--$2\times10^{27}\cms$ from the condition that neutrals must survive ionization while experiencing the amount of heating implied by the \Ha{} narrow component linewidth. The expression derived by \citet{PMBG2006} for the electron diffusion coefficient as a function of the synchrotron X-ray cutoff energy implies that the electron diffusion coefficient for the SNRs studied in their work is within only a factor of a few greater than the Bohm diffusion coefficient.

The spectral shape and the sharply peaked radial profiles of the synchrotron X-ray emission at the forward shocks of young SNRs require a postshock magnetic field strength of the order of $100\muG$ \citep{VL2003,VBK2005,Ballet2006}. Since compression alone is insufficient to produce such a gain in field strength, it is thought that perturbations in the preshock medium generated by the CR streaming instability are non-linearly amplified beyond $\delta{}B/B_0=1$ \citep{BL2001,Bell2004,PMBG2006}. While some simulations show that there exist rapidly growing nonresonant modes \citep{ZPV2008}, these are saturated at $\delta{}B/B_0\sim1$ in other simulations \citep{NPS2008}, and the problem of magnetic field amplification in the CR precursor remains an unresolved. Another mechanism for wave generation is the CR driven acoustic instability \citep{Drury1984,D-F1986,Chalov1988,KJR1992}, which may play a role in prolonging the confinement of high energy CRs in the CR precursor \citep{Berezhko1986,MD2006,DM2007}.

In general, strong MHD turbulence in the CR precursor leads to energy dissipation by wave damping that will affect the structure of the shock \citep{MV1982,CBAV2008}. Although theories of wave damping exist \citep[see e.g.][]{Whang1997}, neither the rate of energy dissipation nor the fractions of the dissipated energy going into internal energy of the various components of the flow (CR electrons and ions, and thermal electrons and ions) are known from observations. In models of DSA, the original treatment of Alfv\'{e}n wave damping in CR modified shocks by \citet{VM1981} is commonly adopted. The damping of acoustic waves, though less common in models of DSA, may also substantially heat the ions \citep{D-F1986} or electrons \citep{GLR2007} of the thermal component. 

A further quantity important for DSA is the efficiency of injection of particles from the thermal population into the CR population. The fraction of swept up thermal protons injected into the acceleration process in SNRs is thought to lie in the range $10^{-4}$--$10^{-2}$ \citep{VBK2003,EC2005} if particle acceleration is efficient.

If the structure of a non-radiative shock is modified by CRs, several subtle signatures in the optical emission from the shock are expected \citep{Raymond2001}. One signature would be a FWHM of the narrow \Ha{} component broader than $20\kms$, the value expected if there is no CR acceleration for an upstream medium at a temperature of $T_0\sim10^{4}\Kv$. Spectra of many Balmer-dominated filaments associated with shocks over a wide range of Mach numbers, show a narrow \Ha{} component with a FWHM in the range $30$--$50\kms$ \citep[see][Table 1]{SGLS2003}, indicative of some common form of preshock heating. Significant bulk acceleration of the upstream flow through the precursor would also be detectable as a Doppler shift of the narrow component centroid with respect to \Ha{} emission from the upstream gas. With the exception of the filament observed by L07, which also concerns this work, this has not been observed yet. Currently, a CR precursor is the favored mechanism for the inferred preshock heating (\citealp{HRB1994,SRL1994,SGLS2003}; L07), but no self-consistent model of such a precursor has been compared with data. Here we show that the precursor structures predicted by two-fluid models of CR modified shocks including particle injection at the subshock and energy transfer due to the acoustic instability are consistent with the above features of \Ha{} spectra.

Recently, L07 obtained high-resolution \Ha{} echelle spectra of an optical filament at knot g in Tycho's SNR, covering the postshock region and the ionization precursor far upstream. Knot g, located in the eastern rim ($\alpha=00\ahour25\aminute56.5\asecond$, $\delta=64^\circ09^\prime28\arcsec$, J2000.0), is the brightest region in \Ha{} emission in the remnant. The synchrotron X-ray emission is also particularly bright in this region \citep{Decourchelle-etal2001,Hwang-etal2002}. Radio data and \HI{} absorption studies suggest that the northeastern rim is decelerating into an inhomogeneous ambient medium, possibly the edges of a molecular cloud \citep[see][and references therein]{LKT2004}. The observations by L07 have spatially resolved a steep \Ha{} (narrow component) flux increase ahead of the shock discontinuity, distinct from the photoionization precursor (G00), which L07 attribute to enhanced emission from a CR precursor. They also reported a broadening of the narrow component linewidth by $15\kms$ and a redward Doppler shift of the narrow component centroid with respect to that of the distant upstream \Ha{} emission in this region of $5\kms$. 

In this paper, we provide a self-consistent CR modified shock model applied to the observational data from L07. We model the shock structure with a time-dependent hydrodynamic two-fluid code. We adjust model parameters to obtain a best fit for the calculated spatial \Ha{} profile to the observed profile. The two-fluid equations along with the free parameters and boundary conditions of the shock model are described in Sect.~\ref{sec:hydro}. The method of calculation for the \Ha{} emissivity is given in Sect.~\ref{sec:halpha}. A time dependent transient solution that provides the best fit to the observed spatial \Ha{} profile is presented in Sect.~\ref{sec:trans}. In Sect.~\ref{sec:steady}, we discuss the solution space of steady CR modified shocks, and propose that the branch of solutions for which the CR acceleration efficiency is low, may explain the line broadening of the narrow component of the \Ha{} line, observed in many SNRs. We discuss our results in Sect.~\ref{sec:discussion}, and conclude the paper in Sect.~\ref{sec:conclusions}.

\section{Two-fluid theory}\label{sec:hydro}
CR modified shocks are sometimes studied with the two-fluid description, first developed by \citet{D-V1981} and \citet{Axford1982}. The CRs are treated as a massless fluid exerting a bulk pressure on the thermal component. The second velocity moment of the CR transport equation provides a conservation equation governing the CR pressure, which is the quantity that directly affects the dynamics of the thermal component. Since we are primarily interested in the backreaction of the CRs on the gas, it is not essential to follow the CR particle distribution in our calculations.

Two-fluid models have been widely implemented in both plane parallel and spherically symmetric geometry \citep[e.g.][]{MAD1984,BC1988,JK1992,DBK1995,ZBPV1996,FKS2000}. The basic theory has been extended to include, e.g, wave dissipation \citep{VDM1984,WFH2007}, oblique shocks \citep{WDV1986,FJR1995}, particle injection \citep{KJ1990b,ZWD1993,KCW1997}, and radiative cooling \citep{Wag2006}. The shock structures obtained with the two-fluid theory agree well with those obtained from kinetic theory and with those obtained with Monte-Carlo simulations \citep{KJ1995,KJ1997}. The computational expense of solving the two-fluid system numerically is far less than that for the latter two methods. 

The approximation of the two-fluid theory consists in the limitation that the adiabatic exponent for the CR component, $\gamc$, cannot be determined self-consistently, and it is therefore usually assigned the constant value of $4/3$, appropriate for a relativistic gas. In principle, the diffusion coefficient for CRs, $\kappa$, is governed by the spectrum of the scattering wavefield, and is thus a function of particle momentum, space, and time. Since two-fluid models lack information about the CR particle spectrum, a momentum-averaged effective diffusion coefficient is commonly used. 
Although one may prescribe the evolution of the closure parameters, $\kappa$ and $\gamc$, \citep[e.g.][]{MDV1990,JK1992,DDV1994} we choose to keep them constant as a first step to model the non-radiative shock at knot g.

The main consequence of the backreaction by CRs on the gas is adiabatic heating and compression in the vicinity of the shock discontinuity where the CR pressure gradient is largest. When a substantial fraction of the total energy has gone into CRs, the overall compression ratio exceeds 4 and approaches 7, and the shock discontinuity may be entirely smoothed out, i.e. the gas subshock disappears. Such solutions are referred to as \lq\lq{}efficient\rq\rq or \lq\lq{}CR dominated\rq\rq{}. Conversely, if only a small fraction of the shock energy goes into CRs, the solution is termed \lq\lq{}inefficient\rq\rq, and the modification of the shock structure is weak. 

In the case of fast shocks, such as those responsible for Balmer-dominated filaments in SNRs, some regions of parameter space permit up to three distinct solutions for the same distant upstream conditions and shock parameters. Some of these solutions do not exist as time-asymptotic steady states, but if exactly two solutions exist, one of them is an efficient solution and the other is an inefficient one. It is important to perform time-dependent runs to determine whether a steady solution exists as a time-asymptotic state. 

We neglect the dynamics of the wavefield responsible for the scattering of the particles, and thus ignore heating of the gas through Alfv\'{e}n wave damping. Instead, we invoke source terms that represent the decay of sound waves generated by the acoustic instability \citep{D-F1986}. We include injection of CR particles at the gas subshock. 

In the remaining parts of this section, we first write down the equations for the two-fluid system and the relevant source terms. We then list the free parameters and boundary conditions for our model.

\subsection{Equations}\label{ssec:equations}
The following equations govern the two-fluid medium consisting of the thermal component and the CR component in plane-parallel symmetry:
\begin{gather}
\frac{\partial{\rho}}{\partial{}t}+\frac{\partial\rho{}u}{\partial{}x}=0\:,\label{eq:fl-mass}\\
\frac{\partial{\rho{}u}}{\partial{}t}+\frac{\partial\rho{}u^2}{\partial{}x}+\frac{\partial{}\PG}{\partial{}x}+{\frac{\partial{}\PC}{\partial{}x}}=0\:,\label{eq:fl-mom}
\end{gather}
\begin{multline}
\frac{\partial}{\partial{}t}{\left(\frac{\rho{}u^2}{2}+\frac{\PG}{\gamg-1}\right)}+\frac{\partial}{\partial{}x}\left(\frac{\rho{}u^3}{2}+\frac{\gamg\PG{}u}{\gamg-1}+{\PC{}u}\right)\\-\PC\frac{\partial{}u}{\partial{}x}=\SG\:,\label{eq:fl-e}
\end{multline}
\begin{gather}
\frac{\partial{}{\PC}}{\partial{}t}+\frac{\partial{}\PC{}u}{\partial{}x}+\left(\gamc-1\right)\PC\frac{\partial{}u}{\partial{}x}-\kappa\frac{\partial{}^2\PC}{\partial{}x^2}=\SC\:.\label{eq:fl-cr}
\end{gather}
$x$ is the spatial coordinate and $t$ is the temporal coordinate. $\rho$, $\PG$, and $T$ denote the mass density, pressure and temperature of the gas, and $\PC$ and $\kappa$ are the CR pressure and the diffusion coefficient, respectively. The two fluids move with bulk velocity $u$. $\gamg$, and $\gamc$ are the adiabatic indices for the gas and of the CRs, and are set to the constant values $\gamg=5/3$ and $\gamc=4/3$. Throughout the paper, subscripts $G$ and $C$ refer to the gas and the CRs, and subscripts 0, 1, and 2 denote a distant upstream value, an immediate presubshock value, and a postshock value, respectively. Equation~(\ref{eq:fl-mass}) expresses mass conservation of the thermal component, and equation~(\ref{eq:fl-mom}) expresses momentum conservation of the thermal component ($\rho{}u$). Equation (\ref{eq:fl-e}) governs the total energy density of the thermal component $\left((1/2)\rho{}u^2+\PG/(\gamg-1)\right)$, and implies that the total energy density of the thermal component is conserved in the absence of CRs. Equation (\ref{eq:fl-cr}) is derived by multiplying the transport equation appropriate for the isotropic part of the CR distribution function by $(4\pi/3)\,p^3\,v$, where $v$ is the CR particle velocity, and integrating that result over all CR particle momenta. 

The source terms in equations~(\ref{eq:fl-e}) and (\ref{eq:fl-cr}) include energy transfer from the CR component to the thermal component due to the acoustic instability in the CR precursor $(\SCa)$, as well as CR injection $(\SCi)$:
\begin{eqnarray}
\SC&=&\SCi+\SCa\:,\\
\SG&=&-\frac{\SC}{\gamc-1}\:.
\end{eqnarray}
Both contributions essentially transfer energy from one component to the other and the total energy of the two-fluid system is always conserved.

The source terms that represent the energy transfer due to the acoustic instability are identical to those used by \citet{WFH2007}:

\begin{equation}\label{eq:sca}
\SCa=\left\{
\begin{array}{lll}
\displaystyle{}-\frac{\gamc-1}{\gamg-1}\PG&\!\!\!\displaystyle\left(\frac{\kappa}{\aG}\left|\frac{\partial{}\PC}{\partial{}x}\right|\frac{1}{\gamc\PC}-1\right)\frac{1}{\tau}\\
&\displaystyle\textrm{if}\quad\frac{\kappa}{\aG}\left|\frac{\partial{}\PC}{\partial{}x}\right|\frac{1}{\gamc\PC}-1>0\:;\\
\displaystyle{}0&\displaystyle\textrm{otherwise.}
\end{array}\right.
\end{equation}




Sound waves are amplified in regions where the relative gradient of the CR pressure, $\gamc\PC\left/\left|\mathrm{d}\PC/\mathrm{d}x\right|\right.$, exceeds a critical length scale $\aG/\kappa$. Here, $\aG=\sqrt{\gamg\PG/\rho}$ is the thermal sound speed. This condition for instability is usually satisfied in the CR precursor because the increase in CR pressure scales as $\kappa/\vs$, and $\vs\gg\aG$. The damping of the sound waves leads to a net energy transfer from the CR component to the thermal component. By invoking the source terms (eq.~\ref{eq:sca}), we bypass the initial perturbations and assume that wave damping occurs at the necessary rate to produce the desired energy transfer. The source term is nonzero if the condition for acoustic instability is satisfied. $\tau$ is a time constant that determines the rate of energy transfer, and the term in brackets ensures that the energy transfer drives the flow towards stability.


We may estimate an order of magnitude value for $\tau$ from dimensional arguments. If we assume a Kolmogorov-type turbulence spectrum with the largest spatial scale set by the width of the CR precursor, $\kappa/\vs$, then the cascade time-scale is approximately 
\begin{multline}
\frac{\kappa}{\vs\va}\approx300\yr\left(\frac{\kappa}{10^{24}\cms}\right)\\\times\left(\frac{\vs}{1000\kms}\right)^{-1}\left(\frac{\va}{10\kms}\right)^{-1}\:,
\end{multline}
where $\va$ is the Alfv\'{e}n speed. This time scale may be comparable to $\tau$ for CR modified shocks in SNRs.

\citet{JK1990} (JK90 hereafter), \citet{ZWD1993}, and \citet{KCW1997} have each adopted a different approach to simulate particle injection in two-fluid models. We follow the method employed by JK90, which is a two-fluid version of that employed by \citet{F-G1987}: 
\begin{equation}
\SCi=\frac{1}{2}\,\epsilon\,\rho_1\,u_1\,(\lambda{}{\aG}_2)^2\,w(x-x_\mathrm{subshock})\label{eq:sc_b}
\end{equation}
The source term represents the injection of CR particles with speed $\lambda{}{\aG}_2$ into the immediate postshock flow, and the corresponding removal of energy from the thermal component. Following \citet{F-G1987} and JK90, we set $\lambda=2$. The rate at which energy is injected is proportional to the mass flux through the shock and the parameter $\epsilon$ determines the strength of injection. 

Injection is zero ahead of the subshock and has a Gaussian dependence on distance, $w$, behind the subshock, which is located at $x_\mathrm{subshock}$. The width of the injection zone must be at least 15 cells. It must also be much narrower than $\kappa/\vs$ in order for the solutions not to be sensitive to the width of the injection region. Following JK90, injection is turned off smoothly but rapidly when the presubshock Mach number with respect to ${\aCG}_1=\sqrt{(\gamg{}{\PG}_1+\gamc{}{\PC}_1)/\rho_1}$ is only slightly larger than unity.

Equations~(\ref{eq:fl-mass})--(\ref{eq:fl-cr}) are solved with a second-order finite-difference Godunov scheme on a uniform cell grid. The CR equation (\ref{eq:fl-cr}) is solved implicitly using a Crank-Nicholson scheme for the diffusion term. The time-step is determined by the Courant-Friedrichs condition with the effective sound speed $\aCG$. The code and technique used are identical to those employed by \citet{WFH2007}.

\subsection{Free parameters}
Let $\phi=\PC/\PG$ be the ratio of the CR pressure to the thermal gas pressure. The free parameters of this model are the shock speed, $\vs$ ($\vs=u_0$ in the shock frame); the ratio of distant upstream CR pressure to distant upstream thermal gas pressure, $\phi_0$; the distant upstream number density, $n_0$, and temperature, $T_0$; the CR diffusion coefficient, $\kappa$; the injection parameter, $\epsilon$; and the time scale of energy transfer due to the acoustic instability, $\tau$. While $\epsilon$, $\tau$, and $\kappa$ are essentially unconstrained free parameters, we may fix or narrow down the ranges of the other parameters from results of previous observations. All the above named parameters remain constant in time, except for $\vs$ which changes as the shock structure evolves.

\subsection{Boundary and initial conditions}
For the results presented in this paper, we have assumed a distance to Tycho's SNR of $2.3\kpc$. The non-radiative shock at knot g is propagating into a photoionization precursor, which is at a temperature of $T_0\approx1.2\times10^4\Kv$, and $n_0\approx1\cmq$ (G00). The shock speed inferred from previous shock modeling is $(\sim2000\pm200)\kms$ (\citealp{CKR1980}; \citealp{SKBW1991}; G01), although these models have been based on the assumption that the available shock energy is converted solely into thermal energy. The inferred shock speed would be higher if a substantial fraction of the shock energy went into the CR component. 

In most environments of the interstellar medium, the CR pressure is comparable to the thermal gas pressure \citep{Ferr1998ApJ}. We have looked at cases for which $1/3\leq\phi_0\leq3$. In general, we find that the consequences of increasing or decreasing $\phi_0$ on the structure and evolution of a shock were very similar to those of increasing or decreasing $\epsilon$. We therefore set $\phi_0=1$.

In the case of knot g, there is some uncertainty in the interpretation of the \Ha{} narrow component linewidth, due to the presence of an intermediate width component (G00; L07). G00, and L07 have found that an adequate fit to the \Ha{} line profile requires three Gaussian components. The component of intermediate width may be produced by protons undergoing secondary charge exchange or by non-thermal motions. Furthermore, the assumption of Gaussian line profiles may not be appropriate \citep{RIL2008}. The presubshock temperature, $T_1$, up to which the flow in the CR precursor is heated, is therefore taken to be an unknown quantity prior to modeling. Our shock model for knot g provides an independent estimate of $T_1$. 

L07 estimated the net acceleration of the flow across the precursor, $\Delta{}u$, to be in the range $60$--$130\kms$. We require this condition to be met in our shock model.

The initial condition $(t=0)$ for a time-dependent run is an ordinary gas shock, for which $\vs=2000\kms$, and for which the CR pressure is homogeneous across the grid at the distant upstream value of $\phi_0=1$. The time to reach a desired shock structure in any time-dependent run should not exceed the age of the remnant, $\tacc=436\yr$. 

We employ free flowing boundary conditions on either side of the grid.

\section{\Ha{} profile calculations}\label{sec:halpha}
We assume that the optical radiation from the shock is not coupled to the hydrodynamics of the flow and calculate the emission from the system as a separate step, using the results from the hydrodynamic simulations as boundary conditions. The assumption of separating the radiative processes from the hydrodynamics in the CR precursor is good because the radiative cooling time is of the order of $10^{10}\,\mathrm{s}$, while the dynamical time-scale across the precursor is of the order of $10^8\,\mathrm{s}$.

In a conventional Balmer-dominated shock, preshock neutral hydrogen atoms swept up by the shock (henceforth referred to as neutrals) are initially unaffected by the collisionless processes that mediate the shock. Within a short distance behind the shock, the neutrals may be collisionally excited before being ionized, giving rise to a narrow \Ha{} line whose width represents the temperature of the \lq\lq{}cold\rq\rq{} preshock neutrals. Preshock protons, on the other hand, are instantaneously heated to a postshock temperature given by the Rankine-Hugoniot jump conditions. Charge transfer between the hot postshock protons and cold neutrals entering the shock gives rise to a population of hot postshock neutrals that are responsible for the broad component of the \Ha{} spectral line profile. A significant fraction of \Ha{} emission comes from \Lyb{} trapping; \Lyb{}  photons from the postshock region are converted to \Ha{} photons through scattering. 

However, this conventional picture of a Balmer-dominated shock does not explicitly account for a CR precursor. The existence of a CR precursor can affect the emitted \Ha{} line in a few ways. On the one hand, neutrals are ionized in the precursor and the number of neutrals reaching the shock front is reduced. On the other hand, collisional excitation of neutrals within the precursor serves as an additional source of \Ha{} (and \Lyb{}) photons. The varying temperature and bulk velocity (and therefore density) of the precursor flow will also affect the rate of \Lyb{} trapping.

We have developed an emission model incorporating the above effects of a CR precursor (Lee et~al. 2008, in preparation), which is briefly described in the following. We consider a flow consisting of hydrogen and helium. Given a temperature and a velocity profile for the precursor, we calculate the spatial evolution of the ionization states of hydrogen and helium in the precursor and postshock regions. The calculations are essentially identical to those of G01, except that we have explicitly included the effects of the precursor with a given temperature and velocity profiles. We include collisional ionization by electrons and protons. We distinguish between fast and slow neutral hydrogen components, and include charge exchange between neutrals and protons. From the resulting spatial profiles of slow hydrogen atoms and fast hydrogen atoms, we calculate the emissivity of the narrow and broad \Ha{} components. For this, we again include excitation by both electrons and protons. Since a significant fraction of \Ha{} emission comes from \Lyb{} trapping, we similarly calculate spatial profiles of the \Lyb{} emissivity and model the radiative transfer of the \Lyb{} line with a Monte Carlo simulation. The flux of \Ha{} photons arising from the radiative transfer of \Lyb{} photons is added to the intrinsic \Ha{} flux to give the total \Ha{} emissivity profile.

We apply our emission model to the hydrodynamic model described in Sect.~\ref{sec:hydro} and compare the results with the observed \Ha{} emissivity profile of L07. For this comparison, the distant upstream H neutral fraction is taken to be 0.85 and He is fully neutral (G00). We assume equal temperatures for the electrons, ions, and neutrals throughout the shock precursor and adopt an equilibration fraction between ion and electron temperatures in the postshock flow of 0.05. We note that results are not very sensitive to the postshock equilibration fraction since excitation by protons in the postshock region is significant for the assumed shock velocity.

\section{Comparison of model calculations with data}\label{sec:trans}
We have performed a thorough investigation of parameter space for the time-dependent two-fluid system described in Sect.~\ref{sec:hydro}. We find that the observed \Ha{} emissivity requires the flow in the CR precursor to be heated to a presubshock temperature of $T_1=10^5\Kv$. However, steady CR modified shock solutions in the high Mach number regime are of two types. The solutions tend to be either smooth, CR dominated solutions or inefficient solutions that are only weakly modified by CRs. In neither branch of solutions are the shock structures very sensitive to the parameters $\vs$, $\phi_0$, $\kappa$, $\tau$, and $\epsilon$. As a consequence we could not find a steady solution in which the gas in the precursor reached a presubshock temperature of $T_1=10^5\Kv$ and was accelerated by $\Delta{}u\approx100\kms$. \citet{BAC2007} and \citet{CBA2008} explored the acceleration time-scale and the evolution of the CR acceleration efficiency for SNRs in the Sedov phase, and they found that the shocks evolve towards a quasi-steady state on a time scale of the order of $10^{3}\yr$. We have, thus, found a transient solution that satisfies the observational constraints. In the following we first present the shock structure of the transient state and then show its subsequent evolution into a CR dominated shock. 
\pagebreak

\subsection{Transient state solution}\label{ssec:trans} 
\begin{figure*}
\centering
\plottwo{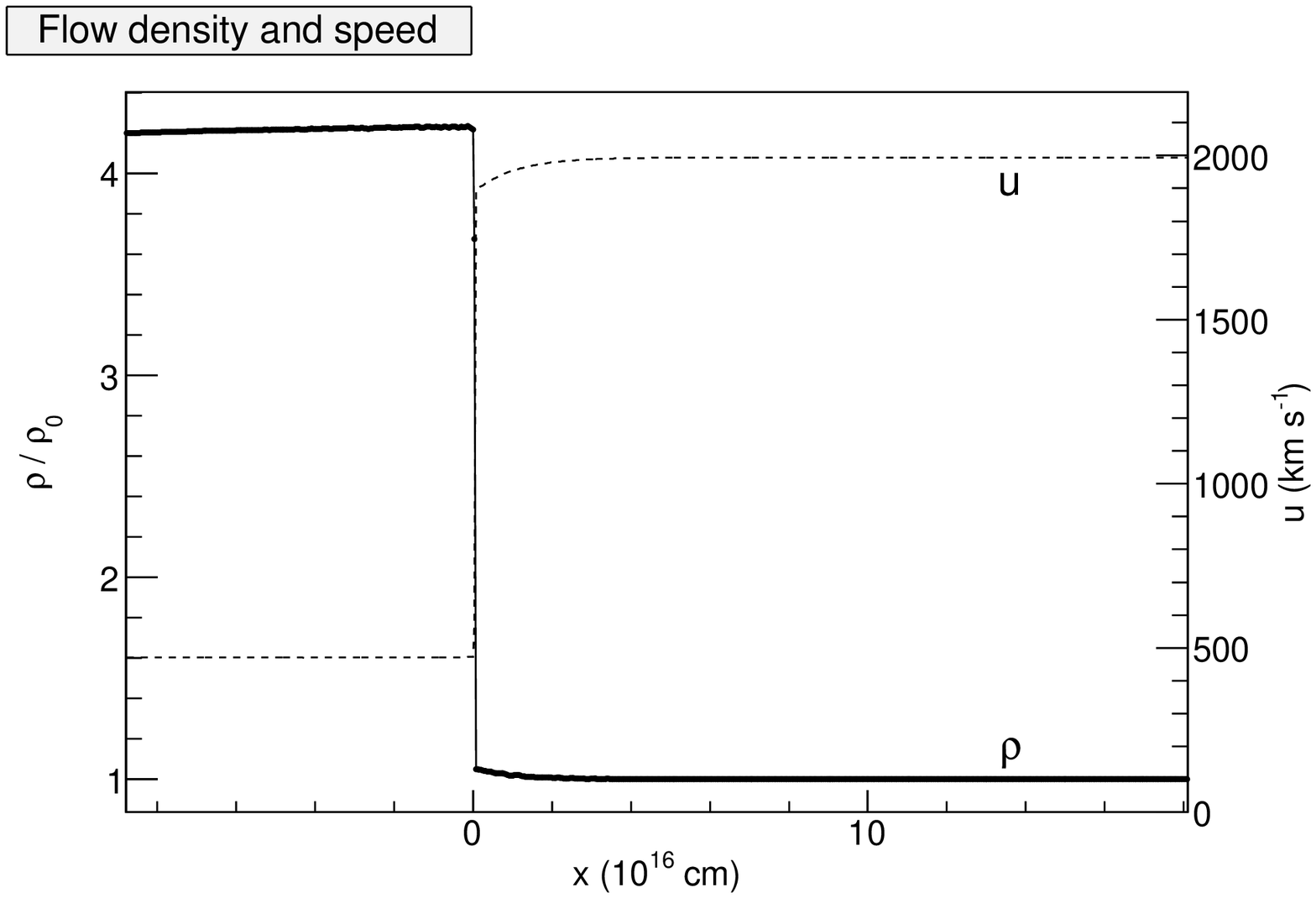}{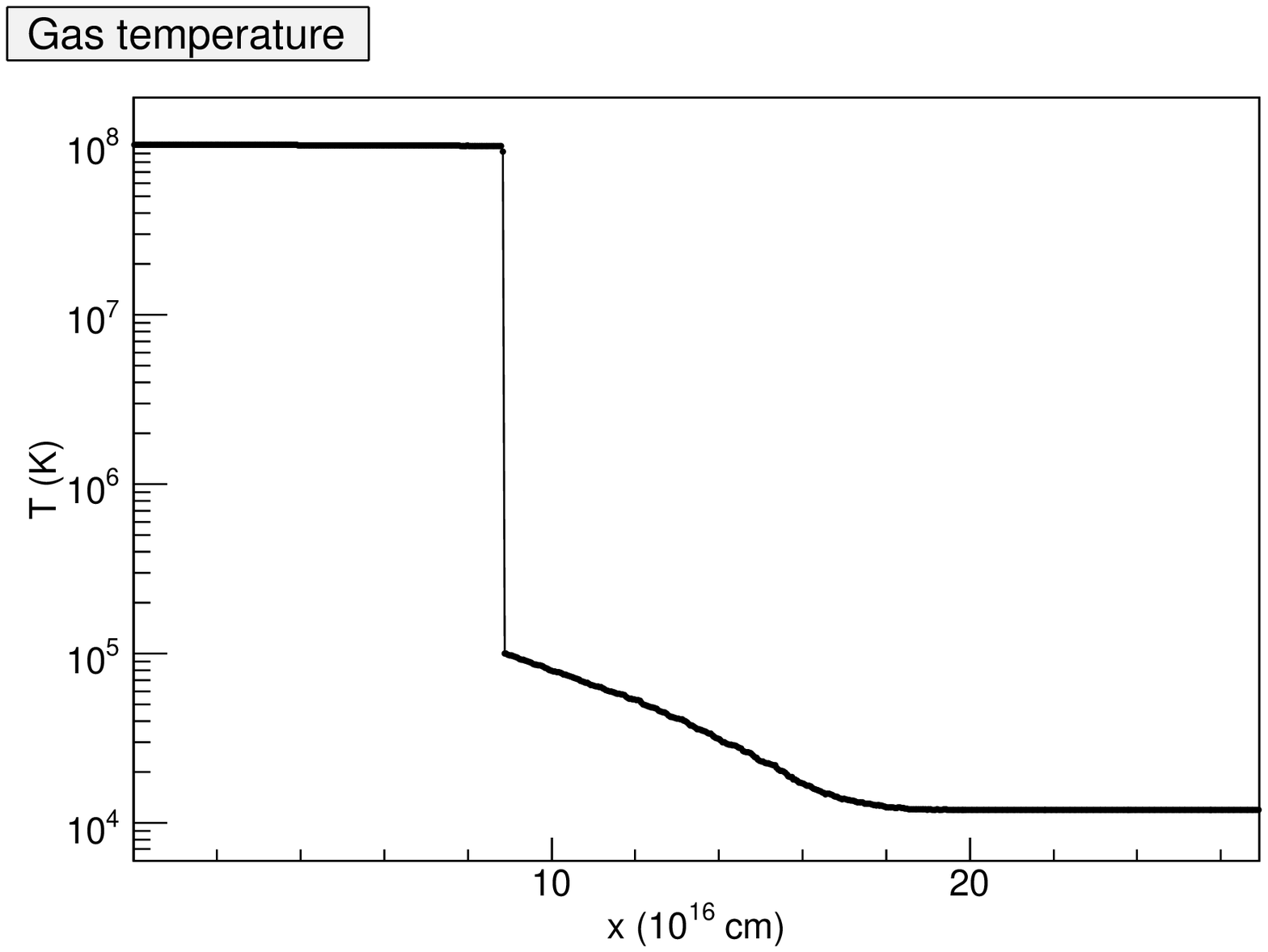}
\plottwo{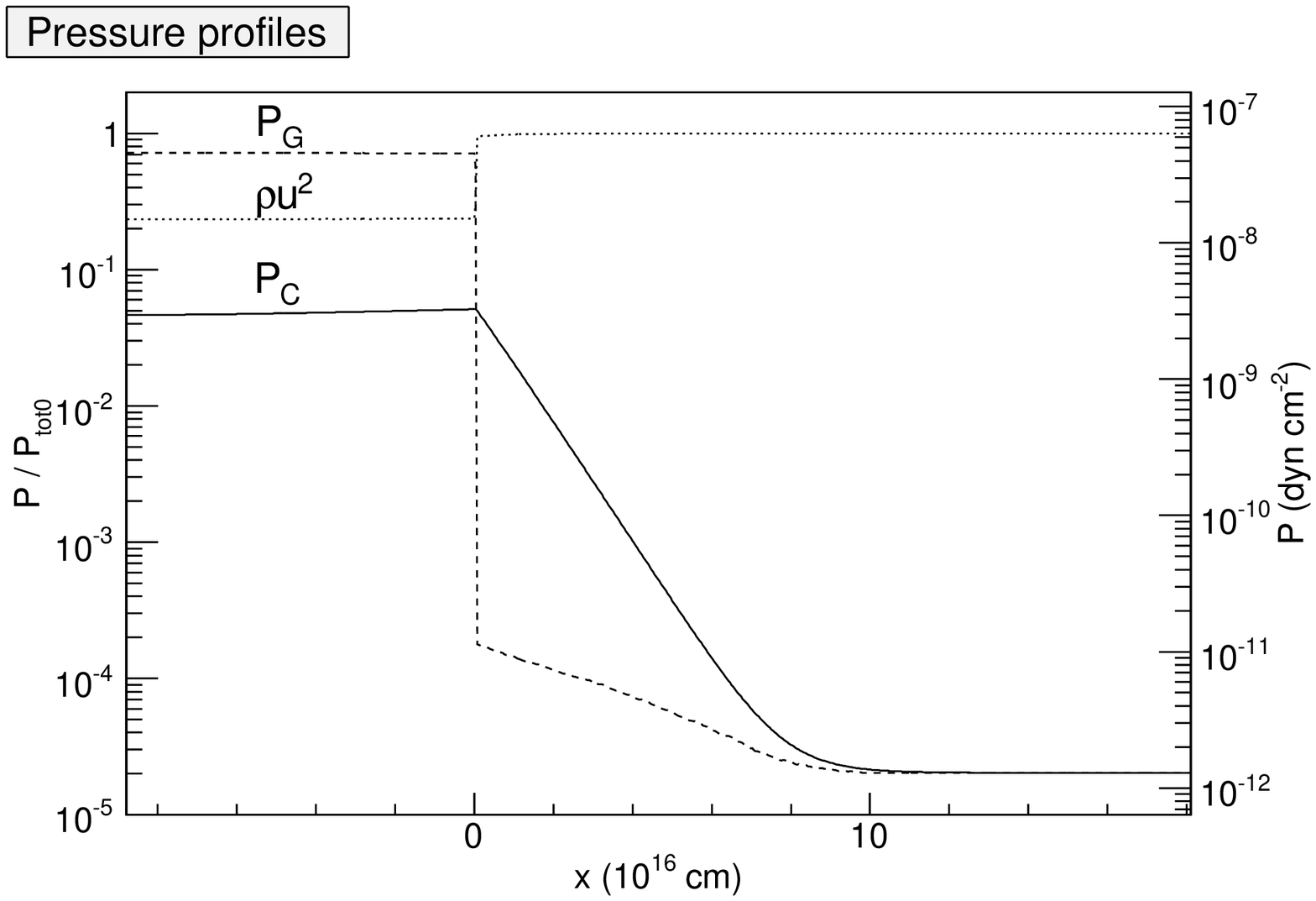}{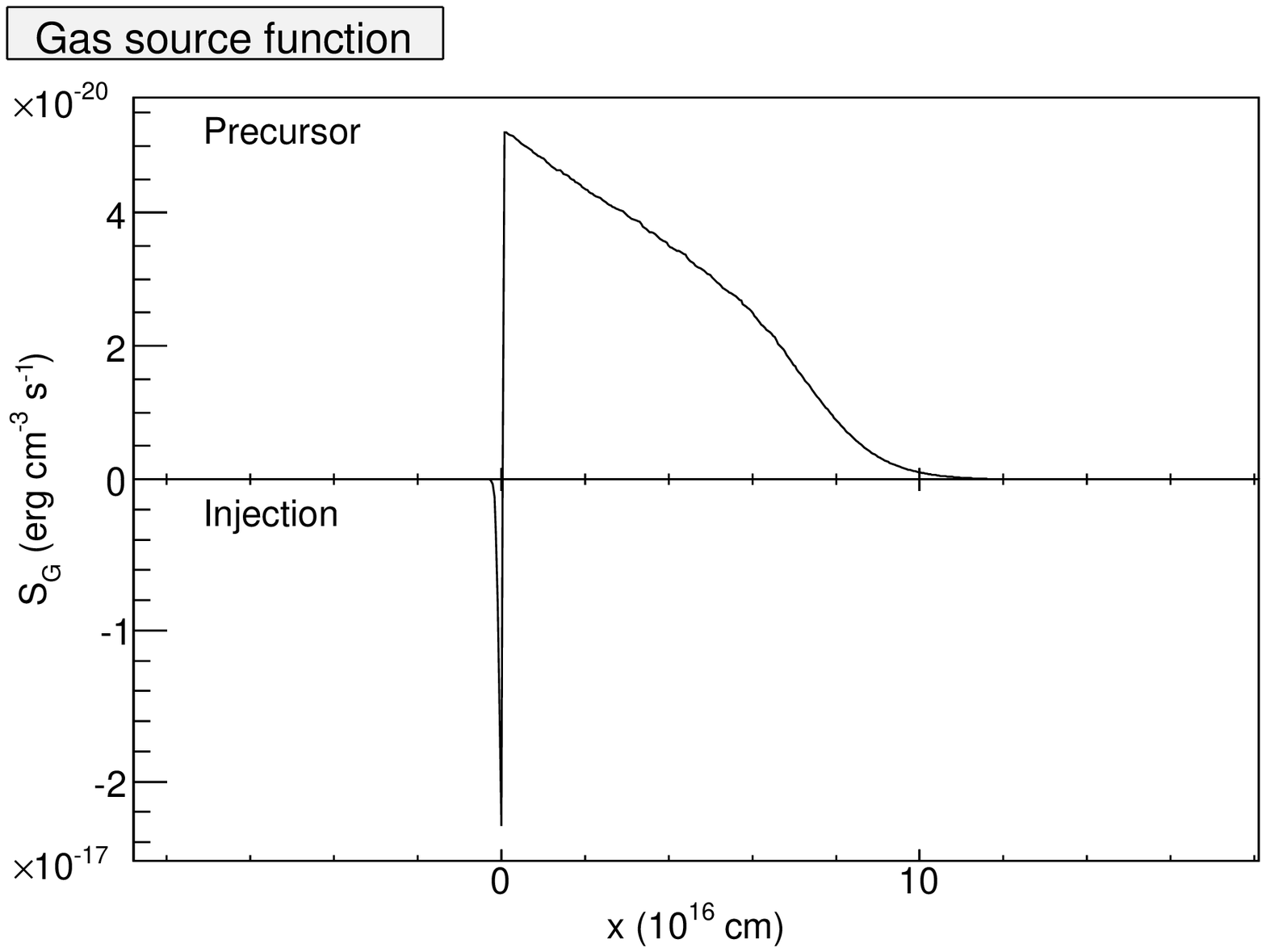}
\plottwo{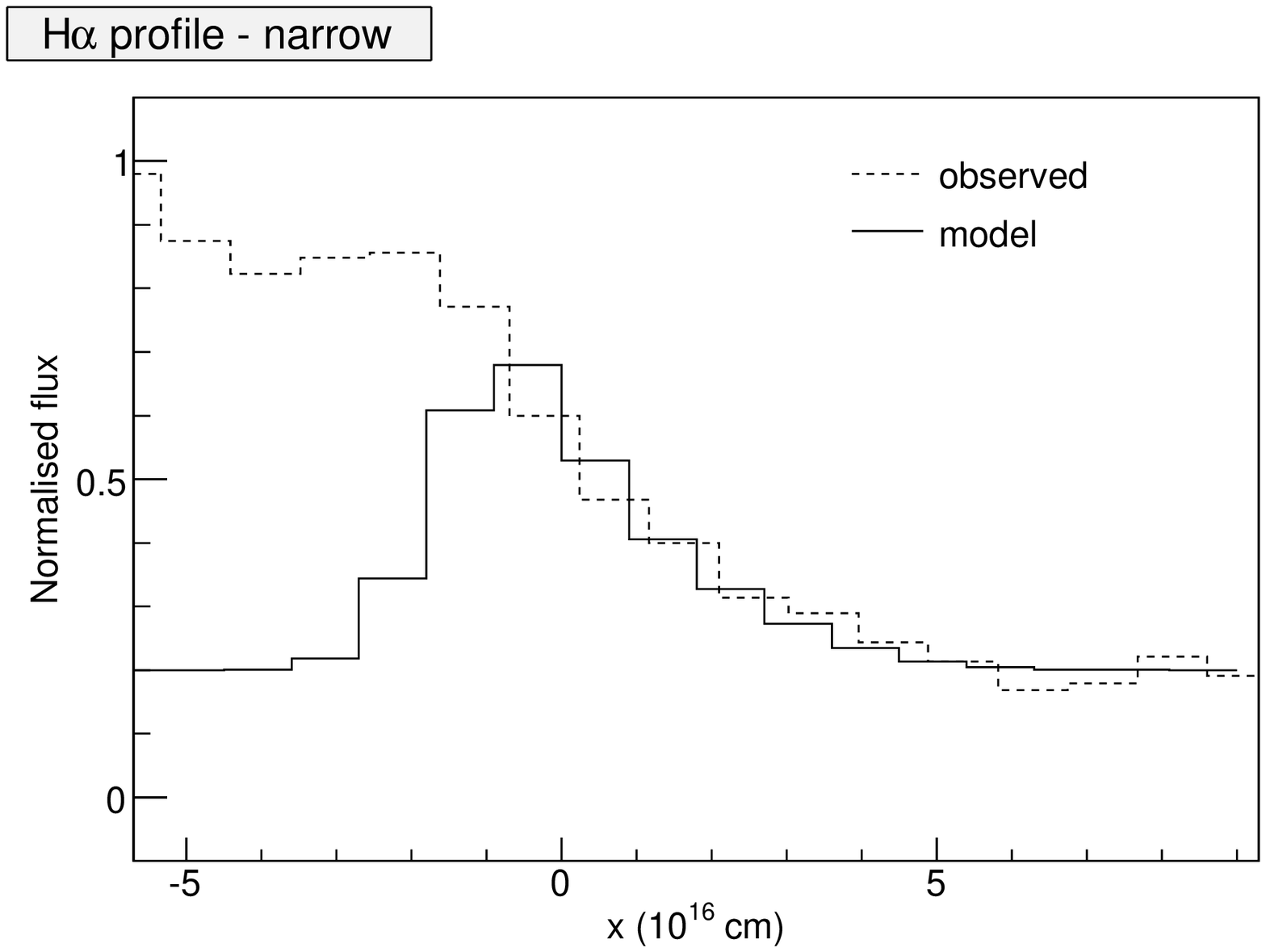}{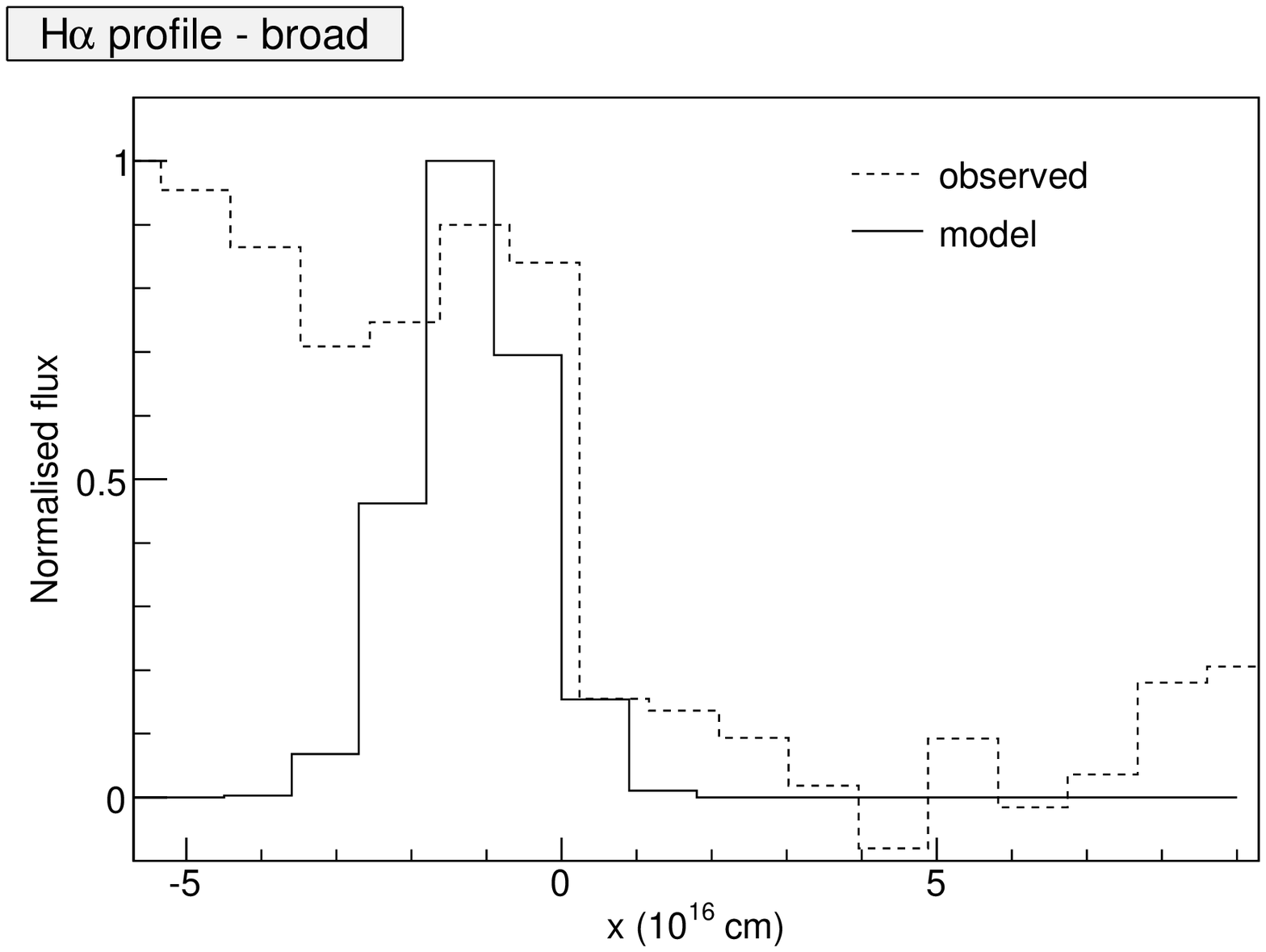}
\caption{Shock structure of the transient solution for which the calculated \Ha{} emissivity profiles match the observed profiles. The gas subshock is located at $x=0$. The transient state is reached at time $t=420\yr$ during the evolution of a shock which was initially $(t=0)$ not modified by CRs (see Fig.~\ref{fig:time}). The shock parameters are, $\vs=2000\kms$, $\phi_0=1$, $\kappa=2\times10^{24}\cms$, $\tau=426\yr$, and $\epsilon=4.2\times10^{-3}$. The shock profiles are shown in the shock frame. The upper part of the panel displaying the source functions is the energy transfer due to the acoustic instability. The lower part of that panel shows the energy transfer due to injection. The model \Ha{} profile is slightly offset along $x$ from the observed \Ha{} profile for visual clarity.}\label{fig:trans}
\end{figure*}

The transient state occurs during the evolution of a shock that is initially not modified by CRs to a shock that is CR dominated. Figure \ref{fig:trans} displays the shock structure of the transient solution for which the calculated \Ha{} emissivity profiles match the observed profiles in the precursor region. In all six panels, the shock profiles are shown in the frame comoving with the shock, and the flow enters the grid from the right.

The location of the gas subshock at $x=0$ is clear from the sudden increase in the broad component flux. We take this to be the outermost non-radiative shock at knot g, or the outermost part of a corrugated shock front (wavy sheet). We are primarily interested in modeling the \Ha{} narrow component flux upstream of the gas subshock. We assume that this emission originates from the upstream gas heated by a CR precursor. The model must also reproduce the immediate postshock narrow and broad component fluxes. The extended emission further downstream is probably due to multiple shock fronts superimposed in the line of sight. We do not require our model to reproduce the extended emission downstream of the immediate postshock region.

This transient state is reached within $420\yr$ of the evolution of a shock for which $\kappa=2\times10^{24}\cms$, $\tau=426\yr$, $\epsilon=4.2\times10^{-3}$. At time $t=0$, the shock is not modified by CRs, $\vs=2000\kms$, and $\phi=1$ throughout the grid. At time $t=420\yr$ (Fig.~\ref{fig:trans}) the shock structure is one in which the thermal pressure still dominates and about $10\%$ of the shock energy has gone into CRs. The bottom two panels show that the narrow and broad \Ha{} emissivity profiles match the observed profiles.

The very close similarity between the values of $\tau$ and $\tacc$ is coincidental, although it does reassure us that the value for $\tau$ is physically reasonable. The value of $\tau$ is also consistent with the estimate of the time scale from dimensional arguments given in Sect.~\ref{ssec:equations}.

The parameters $\kappa$, $\tau$, and $\epsilon$ affect the degree of agreement of the calculated \Ha{} emissivity profile with the observed \Ha{} profile in different ways. The spatial extent of the observed \Ha{} emission fixes the value of $\kappa$ at $2\times10^{24}\cms$. For a given value of $\kappa$, we find that $\tau$ primarily governs the balance between heating and deceleration of the gas in the shock precursor during the evolution of the shock, while $\epsilon$ primarily determines the rate at which the CR acceleration (and shock modification) takes place. 

\begin{figure}
\centering
\includegraphics[width=0.5\textwidth]{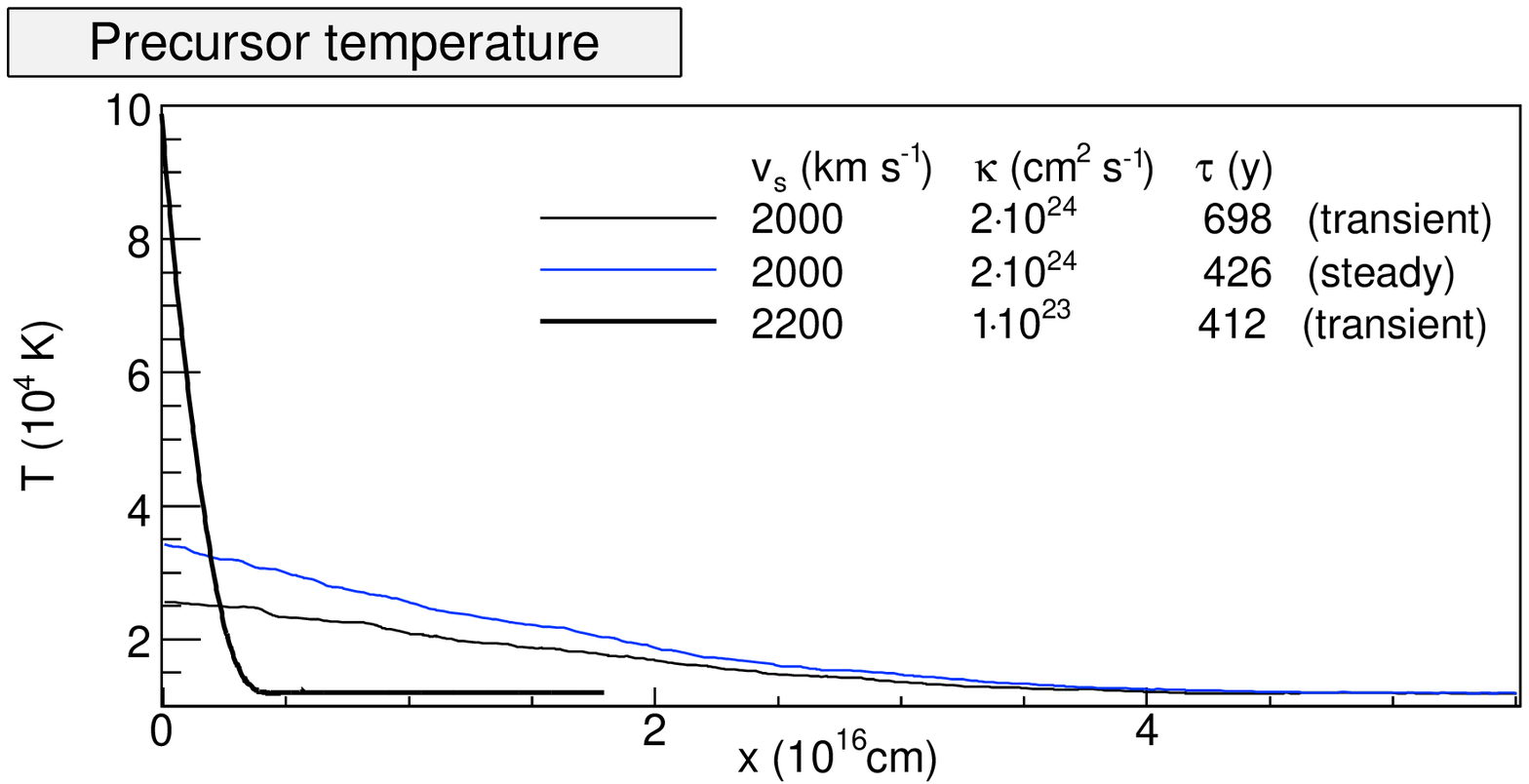}
\includegraphics[width=0.5\textwidth]{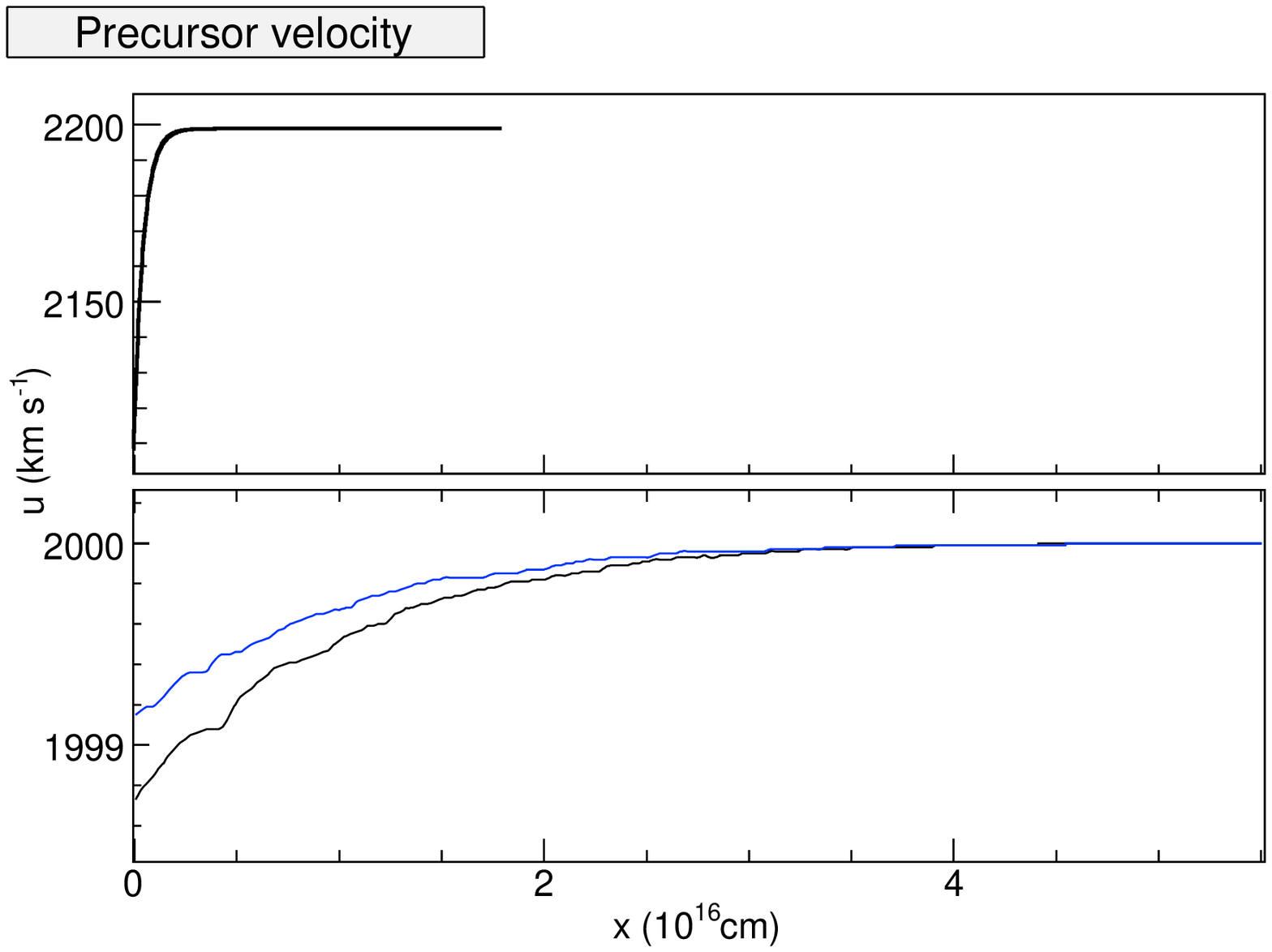}
\includegraphics[width=0.5\textwidth]{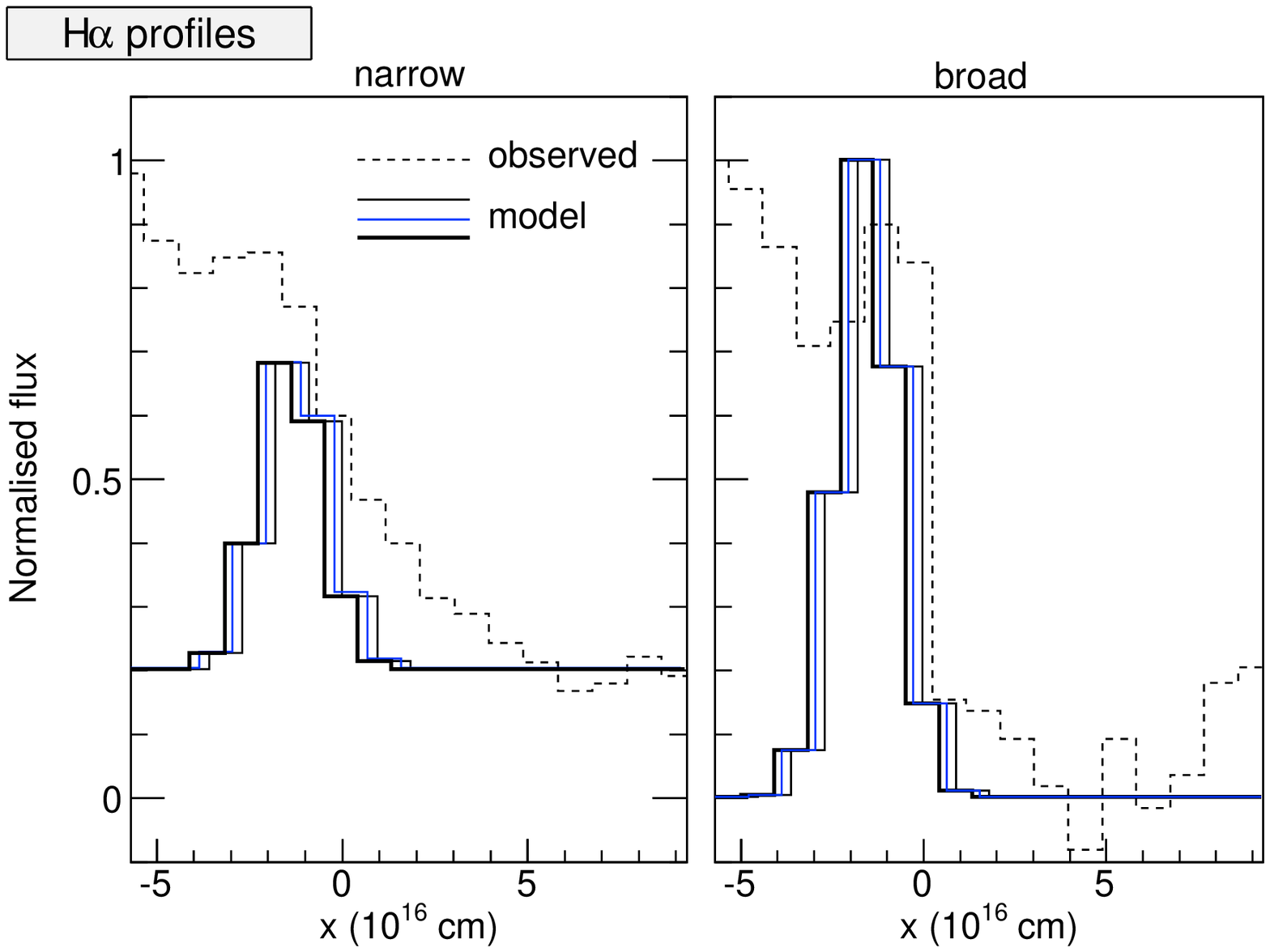}
\caption{Precursor temperature, velocity, and observed and model \Ha{} profiles of shocks that do not include injection. The different values for $\vs$, $\tau$ and $\kappa$ are indicated. The normalisation for the \Ha{} emissivity profiles is arbitrary and separate for the narrow and broad component profiles. The model profiles are slightly offset along $x$ with respect to each other and the observed profile for clarity. The model \Ha{} profiles are very similar, due to the coarse binning and the fact that the emission is dominated by photons from \Lyb{} trapping. None of the shock models without injection give a satisfactory fit to the \Ha{} emissivity data (particularly the narrow component). We conclude that injection is necessary in a model that can explain the data.}\label{fig:badfit}
\end{figure}

Figure \ref{fig:badfit} shows the temperature and velocity structure of the CR precursor for several shocks that do not include injection ($\epsilon=0$). The model for which the values of all parameters other than $\epsilon$ are identical to those of the model shown in Fig.~\ref{fig:trans} does not evolve into a CR dominated shock. The presubshock temperature of the low CR acceleration efficiency steady solution is $T_1=3.4\times10^4\Kv$ and $\Delta{}u<1\kms$. The calculated \Ha{} narrow component profile of this model does not reproduce the observed profile. We find that a lower value of $\tau$ results neither in a larger value of $T_1$ nor a better fit to the data. In the absence of injection, the lowest value of $\tau$ for which the solution evolves into a CR dominated solution is $\tau\approx698\yr$. The preshock heating for this shock as it evolves towards a CR dominated state, is larger than in the case of shocks for which $\tau>698\yr$. However, at $t=420\yr$, this model does not describe the observed \Ha{} either. An alternative to invoking injection in order to decrease the acceleration time-scale is to chose a smaller diffusion coefficient. To obtain a shock structure at $t=420\yr$ for which $T_1=10^5\Kv$ and $\Delta{}u=100\kms$ requires $\vs=2200\kms$, $\kappa=10^{23}\cms$ and $\tau=412\yr$. The calculated \Ha{} profile for this shock model also fails to explain the observed profile. The model \Ha{} profiles in Fig.~\ref{fig:badfit} are very similar, due to the coarse binning and the fact that the emission is dominated by photons from \Lyb{} trapping. The disagreements between the data and the results of the models without injection lead us to conclude that injection is a necessary ingredient in a shock model that adequately describes the emission from knot g.

The shock structure shown in Fig.~\ref{fig:trans} can, in fact, be reached at an earlier time by increasing the value of the injection parameter, $\epsilon$. $\tau$ need only be somewhat reduced, and $\kappa$ remains unchanged. For example, the evolution of a shock that is initially not modified by CRs at $t=0$ will pass through a very similar transient state to that shown in Fig.~\ref{fig:trans} at $t=220\yr$ if $\epsilon=8.0\times10^{-3}$, $\tau=422\yr$, and $\kappa=2\times10^{24}\cms$. The condition that the time at which the transient state is reached must be less than $\tacc$ implies that $\epsilon\approx4.2\times10^{-3}$ is a lower limit for the injection parameter. The transient state cannot be reached with a model in which the only CRs are those swept up by the shock (i.e. $\epsilon=0$), even if $\phi_0=3$.

\subsection{Time evolution of model}\label{ssec:ev}

\begin{figure}
\centering
\includegraphics[width=0.5\textwidth]{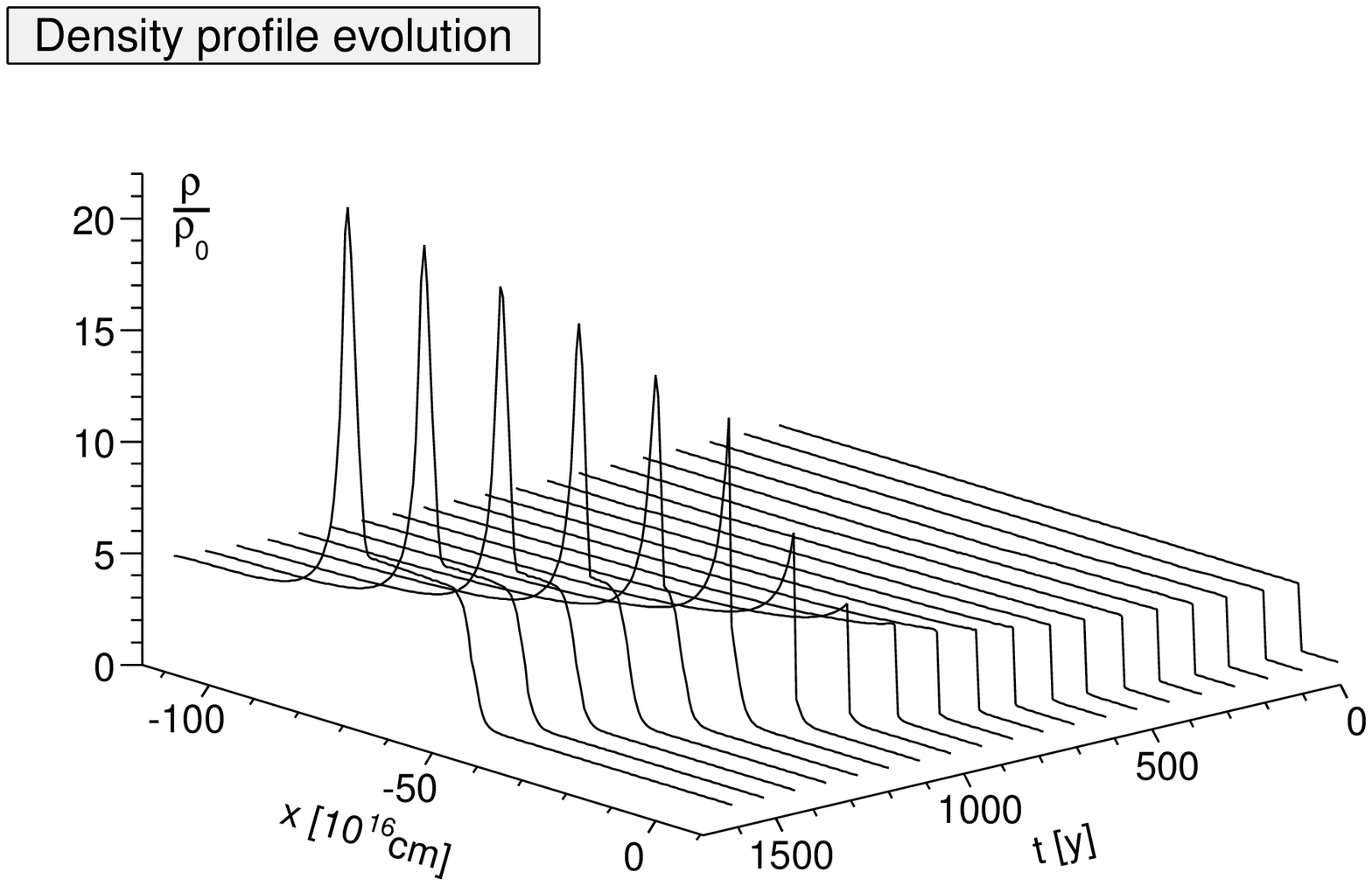}
\includegraphics[width=0.5\textwidth]{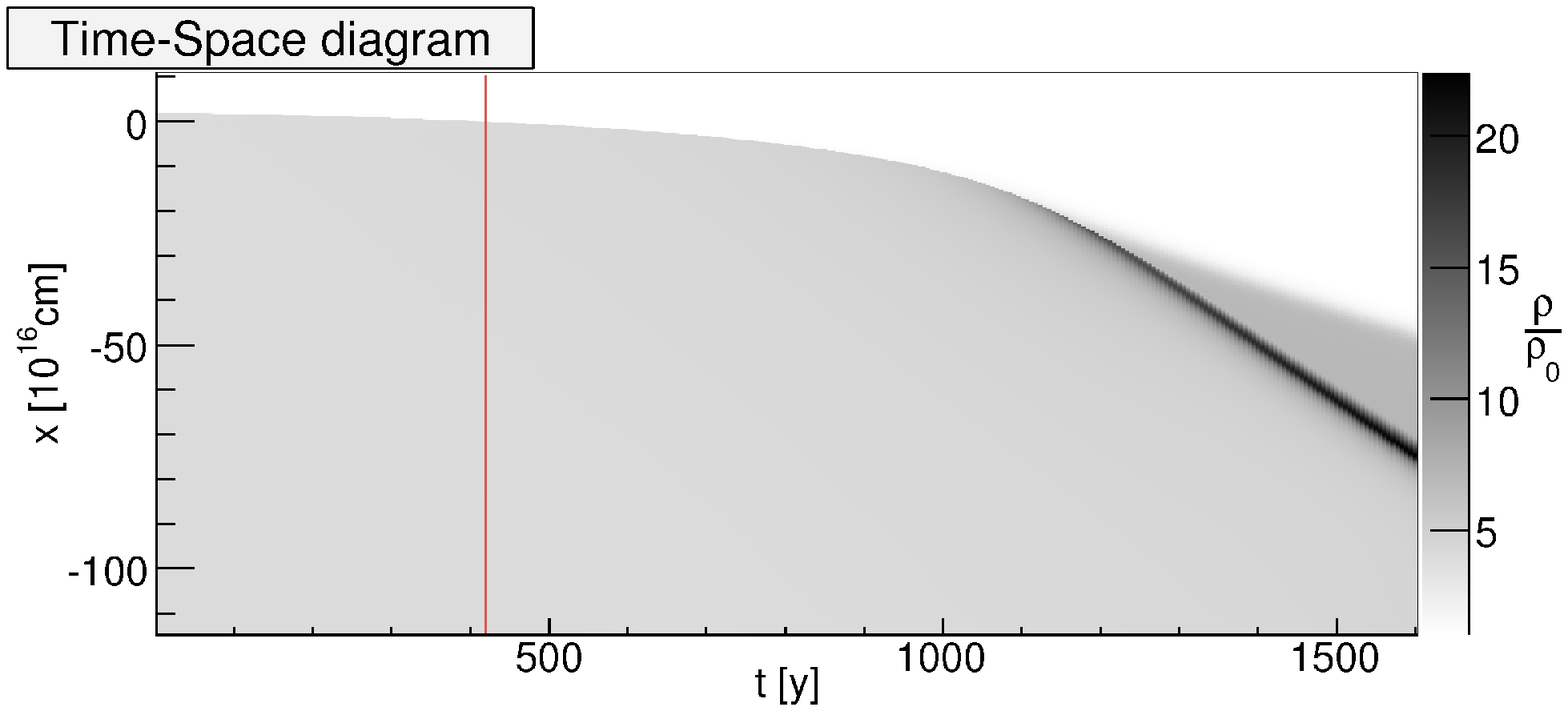}
\includegraphics[width=0.5\textwidth]{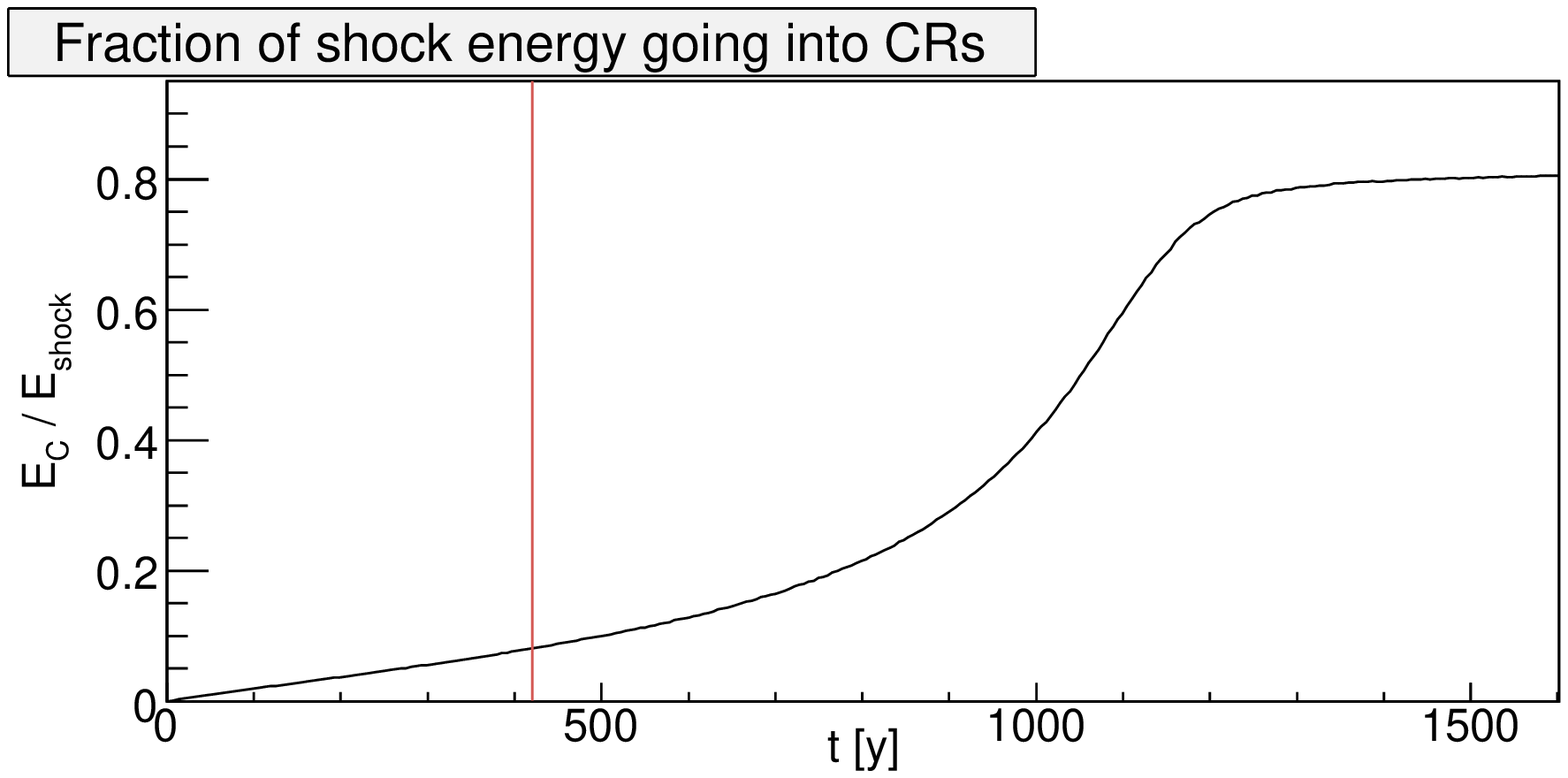}
\caption{Evolution of a shock, which was initially not modified by CRs, and for which $\vs=2000\kms$, $\phi_0=1$, $\kappa=2\times10^{24}\cms$, $\tau=426\yr$, $\epsilon=4.2\times10^{-3}$. The transient state shown in Fig.~\ref{fig:trans} occurs during a phase in which the thermal pressure dominates. During the rapid transition to a CR dominated, smooth shock, beginning at $t\sim1000\yr$, a density spike forms (see Fig.~\ref{fig:spike}). All profiles are shown in the frame comoving with the shock at $t=0$.  The time-space diagram shows the same results as the sequence of profiles. The flow enters the grid from the top. The vertical line in the time-space diagram and in the panel showing the fraction of shock energy going into CRs as a function of time marks the instant in time the transient state occurs ($t=420\yr$).}\label{fig:time}
\end{figure}

\begin{figure}
\centering
\includegraphics[width=0.5\textwidth]{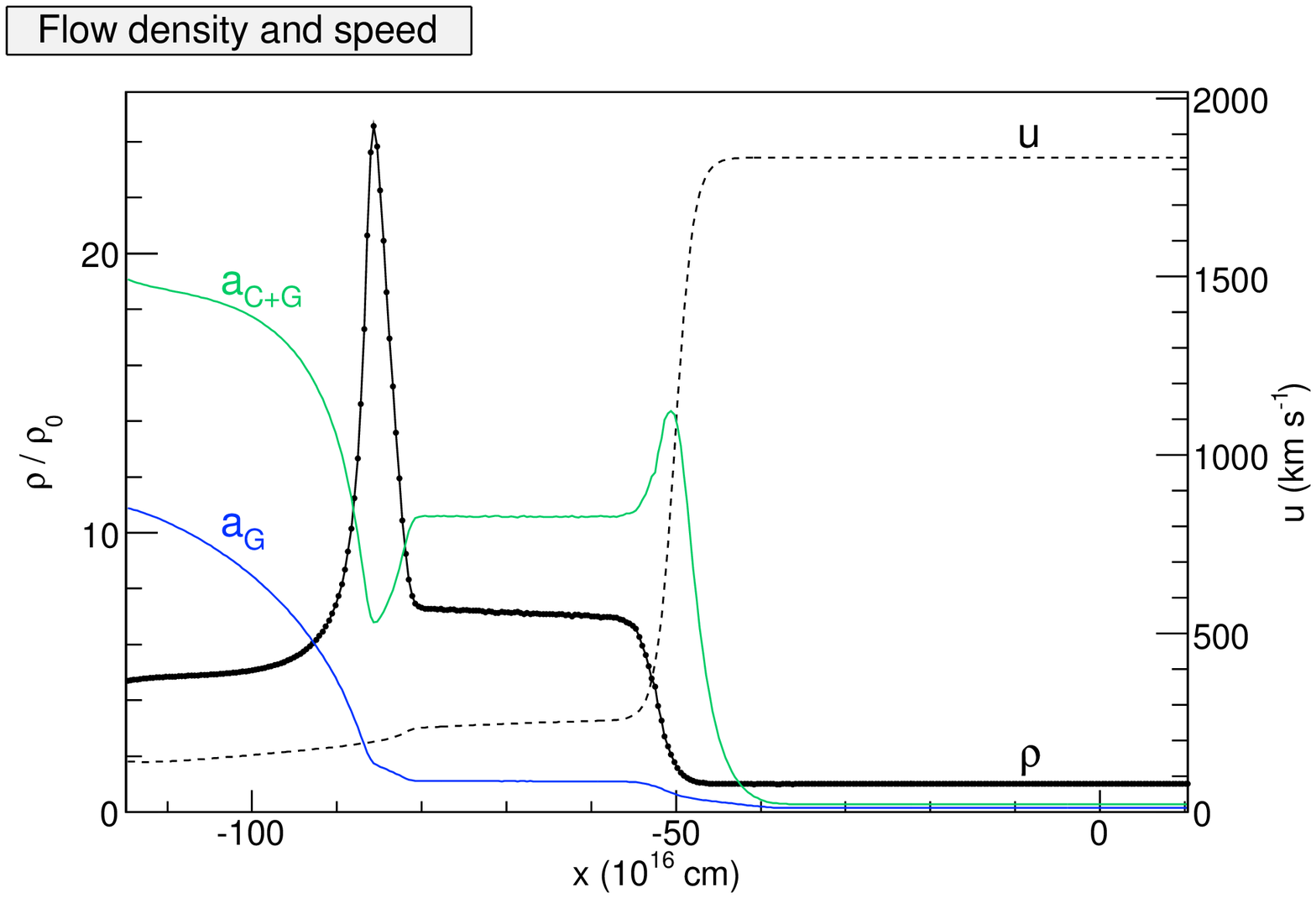}
\includegraphics[width=0.5\textwidth]{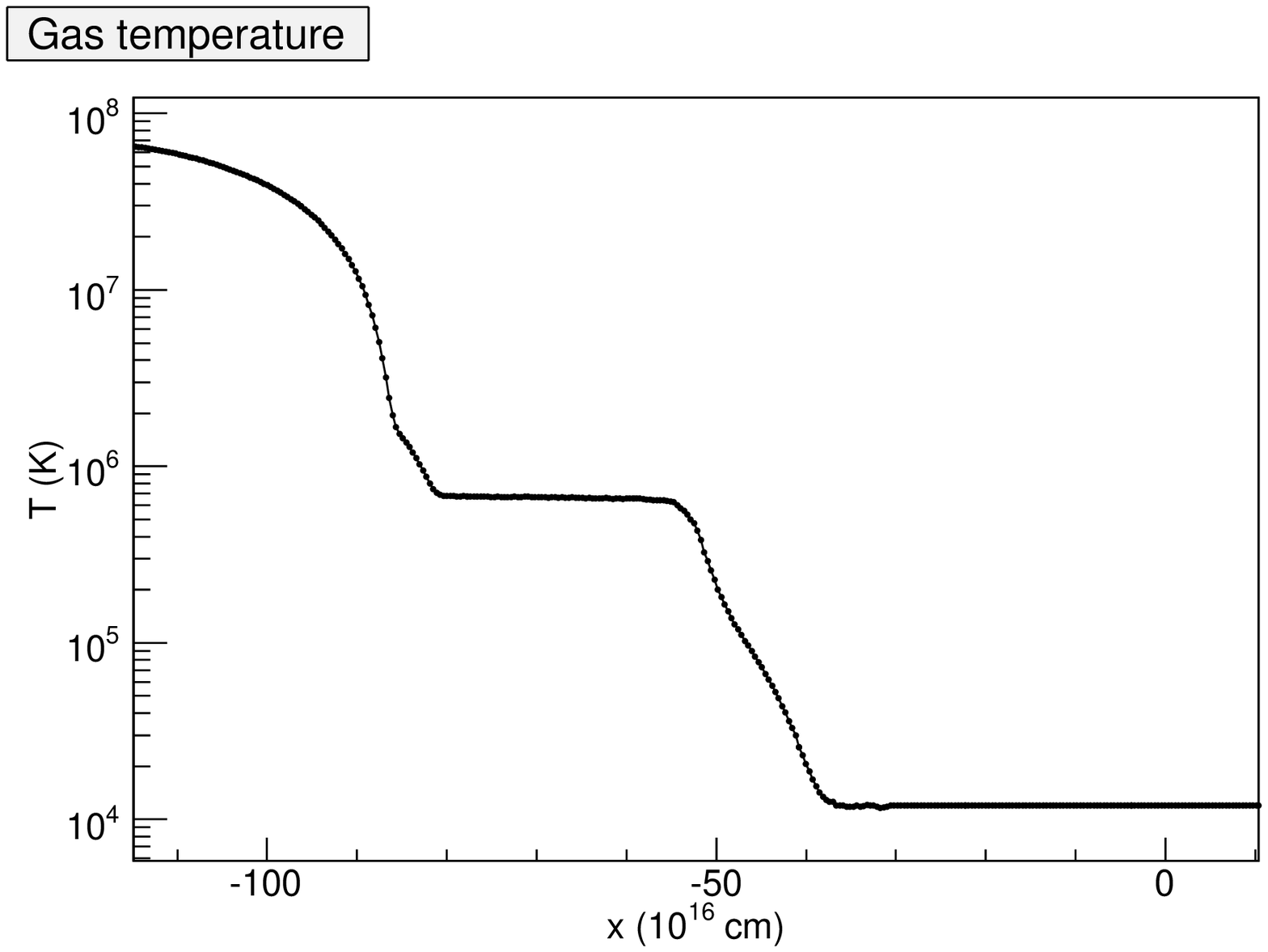}
\includegraphics[width=0.5\textwidth]{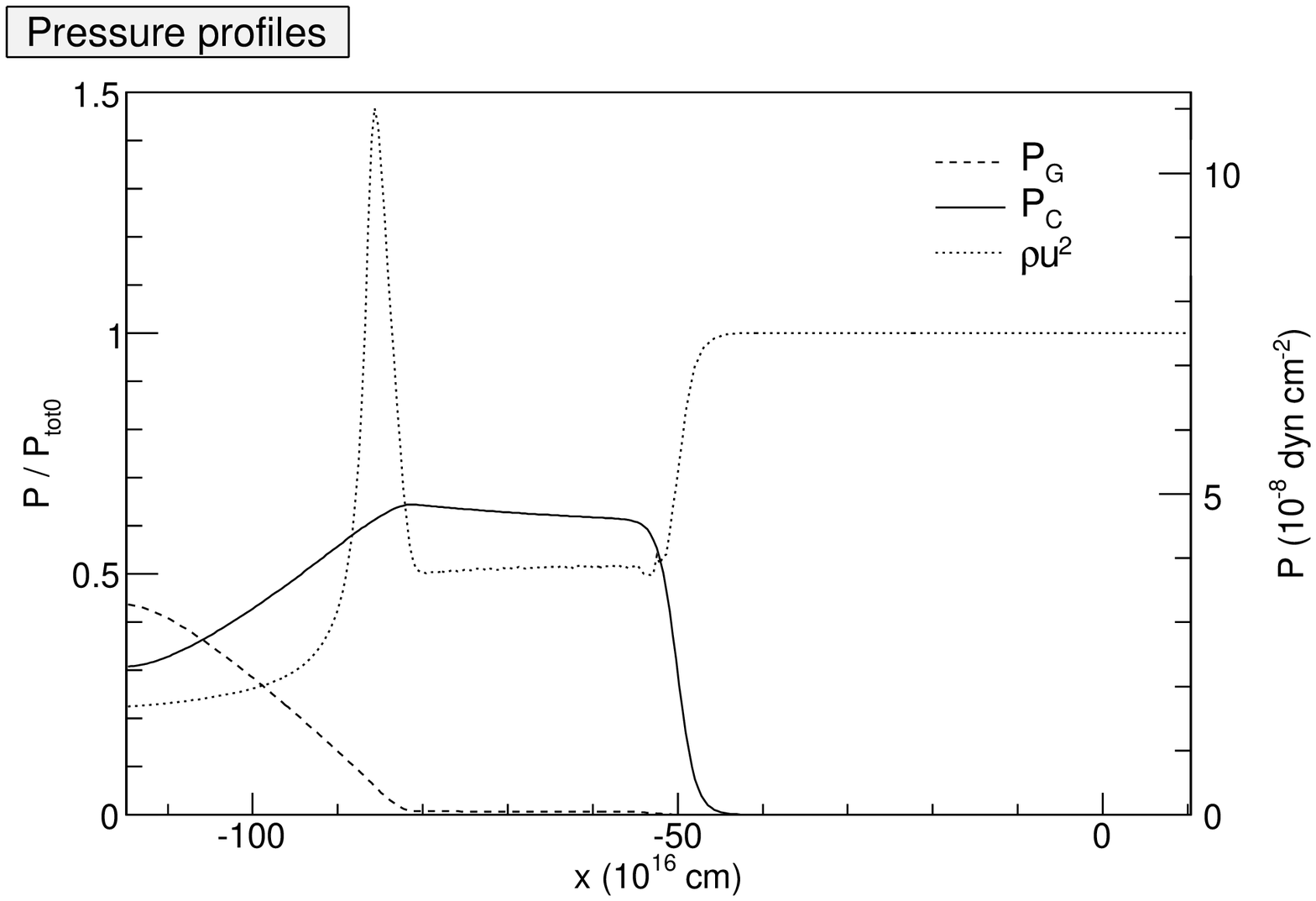}
\caption{Shock structure at time $t=1600\yr$ in the evolution of the model shown in Fig.~\ref{fig:time}. A density spike has formed during the rapid transition to a CR dominated shock and is now traveling downstream. The flow upstream of the density spike is supersonic with respect to $\aG$, but subsonic with respect to $\aCG$, while the flow downstream of the spike is subsonic with respect to $\aG$. The density spike may be long-lived and may play a role in the enhancement of the magnetic field downstream and affect the morphology of synchrotron emission from shocks that accelerate CRs.}\label{fig:spike}
\end{figure}

In Fig.~\ref{fig:time} we show the entire time evolution of a shock that acquires the transient state presented in Sect.~\ref{ssec:trans} at time $t=420\yr$. The fractional gain in CR pressure initially proceeds approximately linearly in time, and the transient state at $t=420\yr$ occurs during a phase of evolution in which the shock modification is still relatively weak. The transition from a weakly CR modified shock that contains a gas subshock to a smooth CR dominated shock is quite rapid. The shock speed recedes to $\vs\approx1830\kms$. During this phase a density \lq\lq{}spike\rq\rq{} develops due to a temporary overcompression arising from the combined compression across the CR precursor and the gas subshock. Within a region of width comparable to the diffusion length-scale, the density is enhanced by a factor greater than $20$ with respect to the distant upstream density, and a factor of $3$ with respect to the density in the immediate postshock region. The density spike travels downstream at approximately $210\kms$ with respect to the smooth CR dominated shock. In Fig.~\ref{fig:spike} we show the density, velocity, temperature and pressure structure of the shock after it has developed into a CR dominated shock ($t=1600\yr$ in the evolution shown in Fig.~\ref{fig:time}). 

The density spike in time-dependent CR modified shocks was first noticed by \citet{Dorfi1984}, and its formation was explained by \citet{Drury1987} and JK90. The portion of the postshock flow in which the CR pressure dominates is bounded by the forward shock and the density spike. The flow upstream of the density spike is supersonic with respect to $\aG$ but subsonic with respect to $\aCG$, and the flow downstream of the density spike is subsonic with respect to $\aG$. The density spike may therefore be long-lived as it moves downstream, although \citet{JJ1997} have demonstrated that the flow through the density spike forms Rayleigh-Taylor instabilities that reduce the density enhancements slightly. Its formation certainly warrants attention in spherically symmetric or 3D simulations of SNR blast waves as it may play a role in the enhancement of magnetic fields and the morphology of synchrotron emission \citep{CHBD2007}. 

A factor of 2--4 enhancement in the thermal X-ray brightness with respect to the local average may be a signature for a density spike travelling downstream of a strongly CR modified shock. This feature would probably be difficult to detect due to projection effects, contamination from ejecta clumping near the forward shock, and low temperatures of the thermal component (see Fig.~\ref{fig:spike}).

\section{Weakly CR-modified steady shocks and the \Ha{} linewidth of Balmer-dominated filaments}\label{sec:steady}
\begin{figure*}
\centering
\plottwo{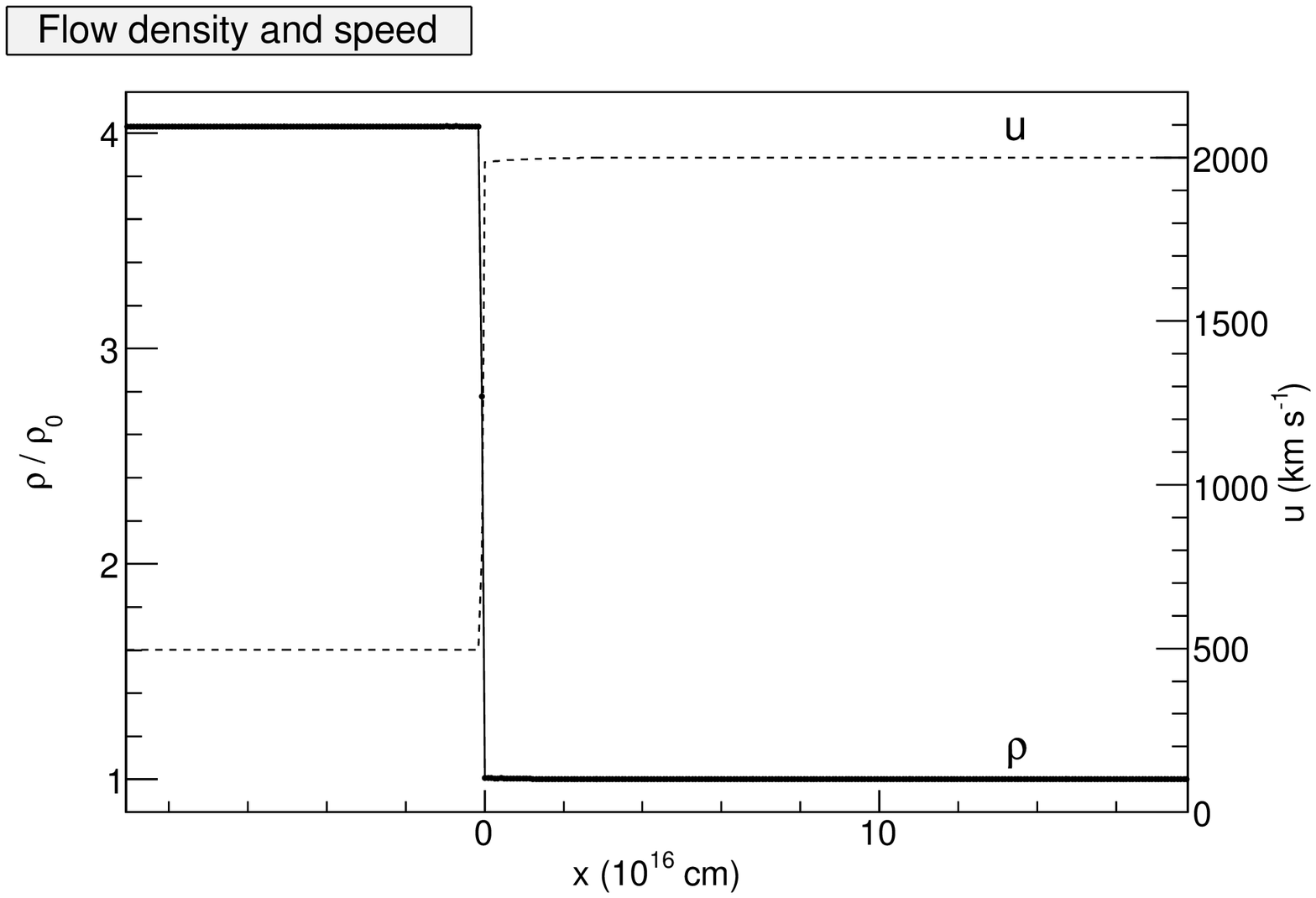}{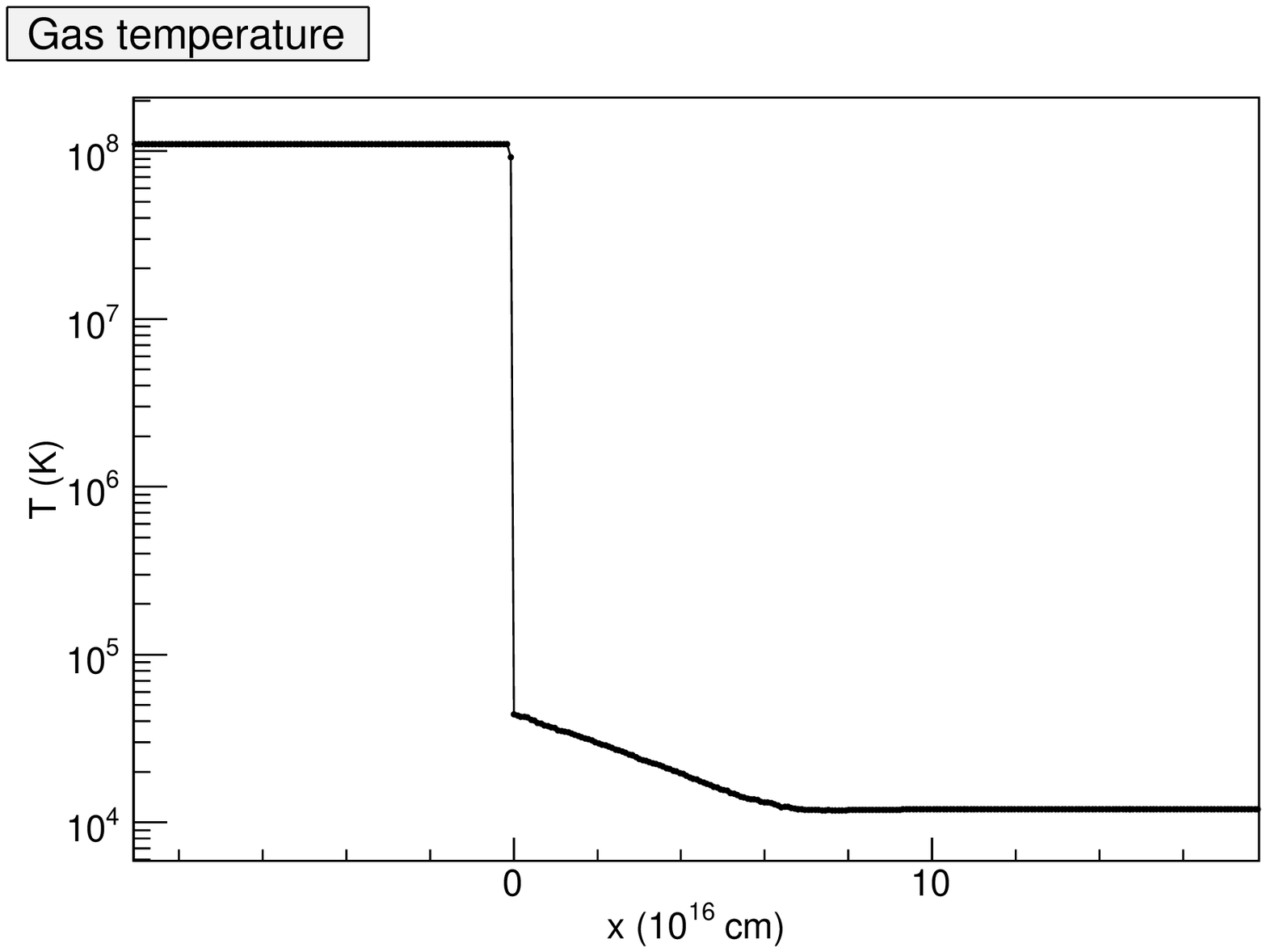}
\plottwo{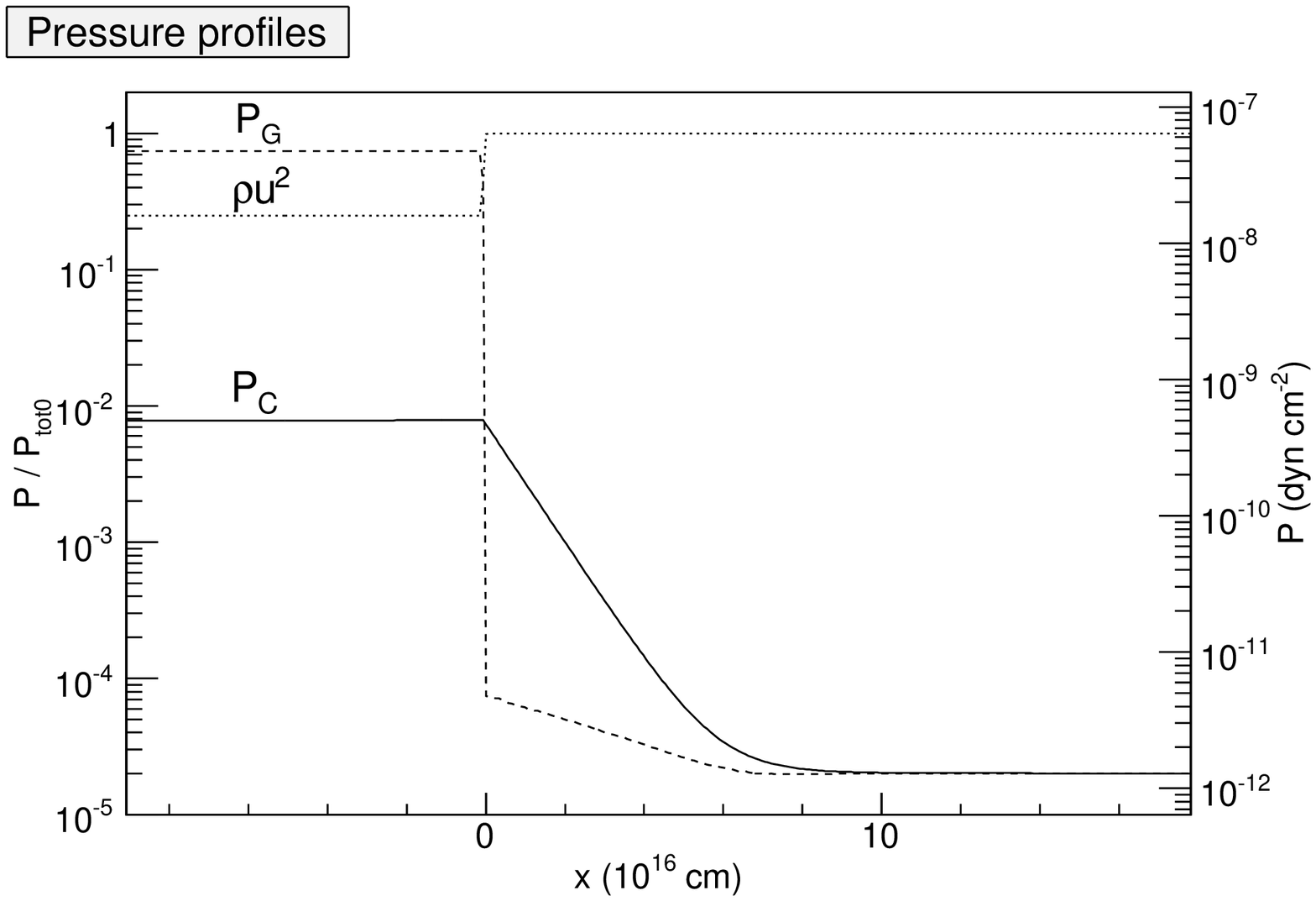}{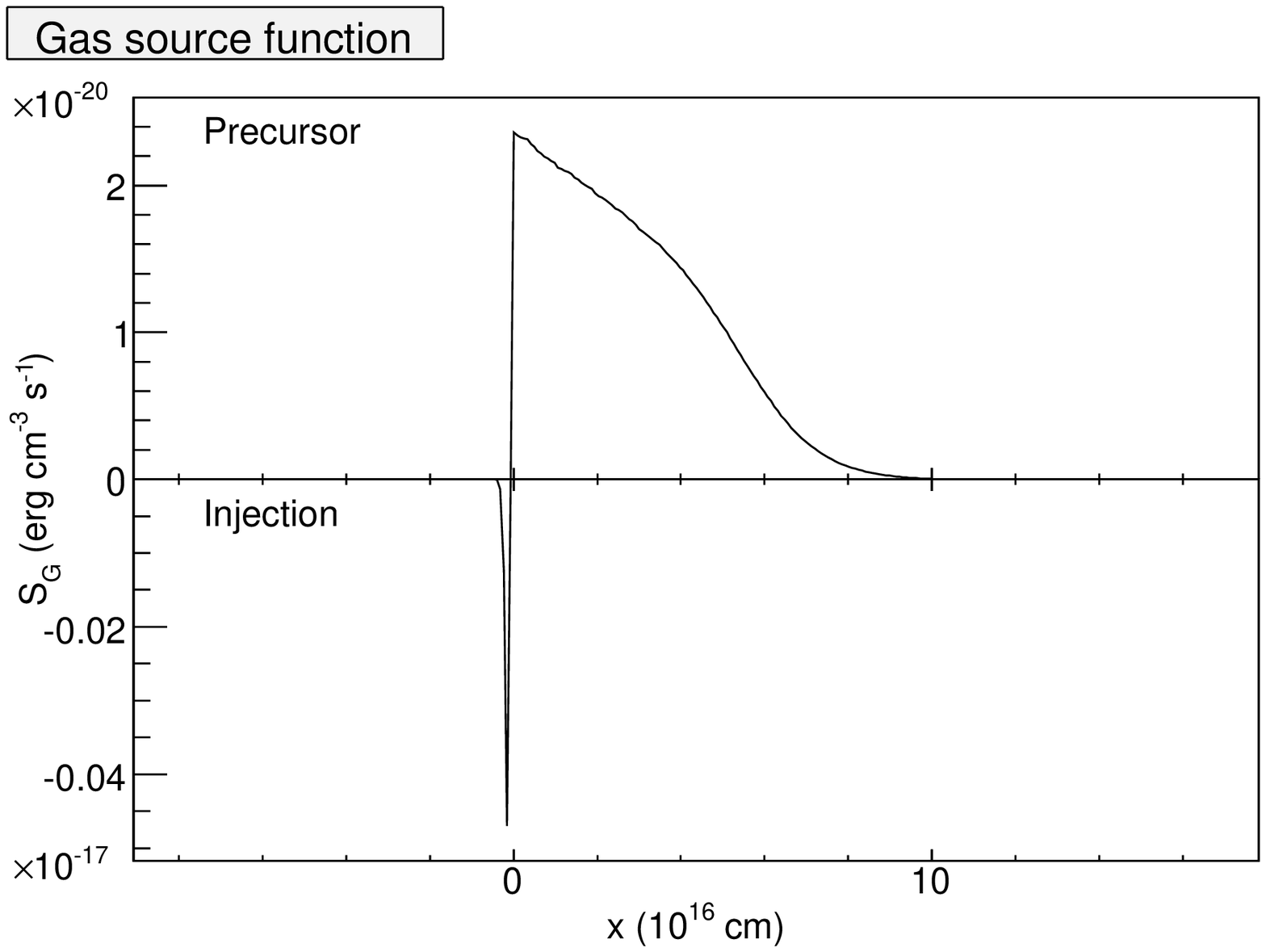}
\caption{Profiles of a steady shock in the low CR acceleration efficiency branch of solutions. $\vs=2000\kms$, $\kappa=2\times10^{24}\cms$, $\phi_0=1$, $\tau=634\yr$, and $\epsilon=1.0\times10^{-5}$. The gas subshock is located at $x=0$, and the profiles are shown in the frame of the shock. The upper part of the panel displaying the source functions is the energy transfer due to the acoustic instability. The lower part of that panel shows the energy transfer due to injection. The set of values for $\tau$ and $\epsilon$ yield the maximum bulk acceleration of the flow through the CR precursor, $\Delta{}u\approx10\kms$. The temperature in the precursor reaches $T_1=4.2\times10^4\Kv$ immediately ahead of the gas subshock. Most solutions in the low CR acceleration efficiency branch are shocks structures similar to the above.}\label{fig:steady}
\end{figure*}

The solution space of two-fluid models for steady, high Mach-number ($M\gtrsim30$), adiabatic, CR modified shocks contains solutions of high CR acceleration efficiency (i.e. $\phi_2\gtrsim1$) and low CR acceleration efficiency (i.e. $\phi_2\ll1$). For some combination of shock parameters and distant upstream conditions there exist multiple solutions. \citet{BK2001} derived exact analytic expressions for the domains in solution space that contain multiple solutions, and the domains in which the solutions exhibit a gas subshock. \citet{MDV2000} have investigated the conditions under which such a bifurcated system may self-regulate.

If $\SG=\SC=0$ and if $\phi_0$ is less than some critical value $\phi_c$ that depends on the shock Mach number, then for any given distant upstream state there are three possible postshock states. Of the three solutions, the one for which the CR acceleration efficiency is intermediate, is unstable \citep{MD1998}, and does not exist as a time-asymptotic state \citep{DZW1994}. The solution for which CR acceleration is most efficient exists if (and only if) it is smooth, i.e. it does not contain a gas subshock. 

Finite source terms determine whether a shock that is accelerating CRs evolves into a CR dominated shock, or a shock for which $\phi_2$ remains small. For high Mach-number shocks $\phi_c\ll1$ if $\SG=\SC=0$. In this case, the solutions bifurcate only if the distant upstream CR pressure is low compared to the pressure of the thermal component, and steady solutions are mostly CR dominated shocks. However, if $\SCa<0$, we find that $\phi_c$ approaches unity, i.e. the solution space is bifurcated if the distant upstream CR pressure is comparable to the thermal gas pressure. This gives rise to the existence of the low CR acceleration efficiency branch in high Mach-number shocks for which $\phi_0\sim\phi_c\sim1$. 

In Fig.~\ref{fig:steady} we show the structure of a shock for which $\vs=2000\kms$, $\kappa=2\times10^{24}\cms$, $\phi_0=1$, $\tau=634\yr$, and $\epsilon=1.0\times10^{-5}$. The pair of values for $\tau$ and $\epsilon$ lie near the upper boundary of values beyond which a low CR acceleration efficiency solution does not exist (see Appendix). The presubshock temperature reaches $T_1=4.2\times10^{4}\Kv$ and the net acceleration across the precursor is $\Delta{}u=10\kms$.  

The shock structures in the low CR acceleration efficiency branch of steady solutions do not vary much with shock speed in the range $300\kms\lesssim\vs\lesssim3000\kms$, which we have explored. We find that in these shocks no more than $1\%$ of the shock energy is channelled into the CR component. The choice of $\tau=634\yr$ and $\epsilon=1.0\times10^{-5}$ (Fig.~\ref{fig:steady}) yield the largest value for $\Delta{}u$ that is possible for a steady shock in the low CR acceleration efficiency branch of solutions, for which $\vs=2000\kms$, $\kappa=2\times10^{24}\cms$, and $\phi_0=1$. 

For very small values of $\tau$ and large values of $\epsilon$, we find solutions in which the presubshock temperature exceeds $10^5\Kv$ and $\Delta{}u\rightarrow0$. In the greater part of parameter space, however, including cases for which $\epsilon=0$, the preshock temperature reaches $2$--$6\times10^4\Kv$, and $\Delta{}u=0$--$10\kms$. We also note that the steady solutions of the low CR acceleration efficiency branch are reached very quickly, usually within $t<100\yr$, if the initial conditions at $t=0$ are those of a shock that is not modified by CRs. 

The insensitive nature of the low CR acceleration efficiency branch of steady solutions to $\tau$, $\epsilon$, and $\vs$, provided that they do not exceed critical values for low CR acceleration efficiency steady solutions to exist, may be an explanation for the small range of FWHM $(30$--$50\kms)$ for the narrow component of the \Ha{} line observed in Balmer-dominated filaments of many SNRs \citep{SGLS2003}. Currently, other models of non-radiative shocks which do not include CRs are not able to predict the FWHM of $30$--$50\kms$. Furthermore, the lack of observed bulk Doppler shift in the narrow component is also consistent with the small values for $\Delta{}u$ that we obtain. Since only a small fraction of the shock energy goes into the CR component, the shock speeds inferred from previous models applied to non-radiative shocks in SNRs remain valid.  

\section{Discussion}\label{sec:discussion}
The results from the time-dependent solutions presented in Sect.~\ref{sec:trans} depend on the assumed distance to Tycho's SNR. The canonical value, also assumed here, is $d=2.3\kpc$. \citet{LKT2004} have briefly reviewed the debate on the distance to Tycho's SNR. While most authors adopt $d=2.3\kpc$ as established by \citet{CKR1980} and subsequently confirmed by several other studies \citep[see][]{Strom1988}, \citet{Schwarz-etal1995} argue for a value of $4.6\pm0.5\kpc$, based on $21\cm$ absorption features. However, \citet{BR1984} found that absorption features at those velocities were present in the spectra of stars within $2.5\kpc$ that were close to Tycho on the sky. The calculations by \citet{VBK2007} together with the high energy $\gamma$-ray flux upper limit from \textit{HEGRA} \citep{HEGRA2001} imply a distance $4\kpc\gtrsim{}d\gtrsim3.3\kpc$. If the distance to the remnant is greater than $2.3\kpc$, then the shock speed inferred from proper motions, and the intrinsic  width of the CR precursor would be larger. The values for $\kappa$ and $\tau$ must be scaled accordingly to obtain a model whose calculated spatial \Ha{} emissivity profile matches the observed profile. $\epsilon$ must also be increased to allow for the desired transient state to be reached within $t<\tacc$. However, \citet{AHMR2008} have applied their improved shock models to the \Ha{} data and obtained a 15\%{} slower shock speed of $\vs=1600$--$1700\kms$ under the assumption that $\PC/\PG=0$. The slower shock speed would lead to a correspondingly smaller distance to Tycho's SNR. Our result that $\sim10\%$ of the shock energy goes into CRs during the current state of the shock at knot g would imply a $5\%$ increase in $\vs$, partially compensating for the lower shock speed inferred by \citet{AHMR2008}.

Although the spatial emissivity profile of the \Ha{} narrow component places a stringent constraint on $\kappa$, a separate upper limit for the width of the CR precursor comes from the fact that a substantial fraction of hydrogen atoms must avoid ionization ahead of the gas subshock in order to produce the observed broad component. Thus, it is important that our model describes the narrow component flux from the CR precursor as well as the immediate postshock broad to narrow component flux ratio. For the values of the diffusion coefficient, distant upstream density, and required preshock temperatures in this problem, the CR precursor length-scale is much shorter than the preshock electron, proton, and photon ionization length-scales, and the H ionization fraction remains nearly constant throughout the CR precursor. This justifies our choice to adopt the distant upstream neutral fraction of $85\%$ derived from the photoionization model by G00 as a fixed boundary condition. A different choice of upstream ionization fraction would affect the immediate postshock ratio of broad to narrow component fluxes, as well as the rate of \Lyb{} trapping in the CR precursor and, therefore, the spatial \Ha{} profile in front of the gas subshock. The upstream density $n_0$ only plays a small role in determining the \Ha{} narrow component flux from the CR precursor through enhancing \Lyb{} trapping in the precursor region close to the gas subshock. The more significant property that determines the spatial flux profiles is the shock temperature profile, in particular the value of the presubshock temperature.

The presubshock temperature $T_1=10^5\Kv$ inferred from our model is larger than the temperature of $\sim4\times10^4\Kv$ implied by the narrow component linewidth in the data of L07. A possible reason for this is the single-temperature approximation we use for the precursor gas. The neutrals may have a lower temperature if the length scale for charge exchange is a significant fraction of the CR precursor length-scale. This may be the case for shocks propagating at speeds $\vs\gtrsim2000\kms$ into a medium of low fractional ionization.

It should be noted that the calculated \Ha{} emissivity is sensitive to the ratio of electron to ion temperatures in the CR precursor. We have adopted a ratio of unity, with the assumption that some plasma-physical process provides the requisite electron heating throughout the CR precursor. Possible processes include the resonant exchange of energy between electrons and protons via lower hybrid waves excited, for example, by the two-stream instability due to shock-reflected ions \citep{Laming2001a}, or by the same mechanism that induces high frequency magnetosonic waves \citep{GLR2007}. 

In the mechanism for wave dissipation proposed by \citet{GLR2007} the electrons may attain a temperature of up to $0.3\keV$ in a CR precursor. For a shock for which $\vs=2000\kms$ and the preshock density is $n_0=1\cmq$, this would imply a precursor width less than $\sim10^{16}\cm$ to avoid complete ionization of the neutrals. This is smaller than the precursor width observed in knot g, suggesting that the electron temperature is considerably lower.

Based on a study of the Tycho remnant's X-ray morphology and spectral characteristics, \citet{Warren-etal2005} proposed that the forward shock at most azimuthal angles is strongly CR modified. A shock compression ratio approaching 7 could explain the proximity between contact discontinuity and forward shock, though projection effects may allow a smaller compression ratio \citep{Cassam-Chenai-etal2008}. Knot g is situated at an azimuthal angle approximately $80^\circ$ east of north, and is clearly recognizable as a local minimum in the ratio of the radius of the contact discontinuity to that of the forward shock \citep[Fig.~4 of][]{Warren-etal2005}. Assuming that the X-ray shock and the Balmer-dominated shock at knot g are directly associated, the small ratio of radii in the direction of knot g is consistent with our shock model in the transient state: the shock in the transient state is still in the phase in which the shock modification due to CRs is relatively weak, although it is developing into a CR dominated shock. 

Although the evolution of the shock in our time-dependent run into a CR dominated state (Sect.~\ref{ssec:ev}) is physical within the framework of the two-fluid theory (Sect.~\ref{sec:hydro}), the question of whether it is a realistic description of the subsequent evolution of the Balmer-dominated shock at knot g must be approached with care. Our plane-parallel model accounts neither for adiabatic losses due to expansions perpendicular to the shock normal, nor for particle escape upstream. These two effects may reduce the CR acceleration efficiency appreciably. For example, the steady state, kinetic shock models of \citet{CBA2008} that include particle escape, and the spherically symmetric, kinetic models of \citet{KJ2006} predict that the acceleration efficiency, which they define as ${\PC}_2/\rho_0u_0$, reaches $60\%$ during the Sedov phase, whereas we obtain an equivalent efficiency greater than $70\%$ in our time-dependent run. The absence of losses in our model should not significantly affect the early phase of the evolution during which CR modification is still weak, and during which the transient state that describes the Balmer-dominated shock at knot g occurs.


The shock models presented in this work are based on the interpretation of the observations by L07 that the steep rise in the flux of the \Ha{} narrow component ahead of the shock front is due to a CR precursor. The possibility of other types of precursors remains. Of these, a fast neutral precursor is currently thought to be the most likely alternative candidate. A fast neutral precursor is mediated by the hot postshock neutrals (those responsible for the broad component of the \Ha{} line), which escape upstream and deposit some energy via charge exchange and elastic collisions. The calculations by \citet{LR1995} and \citet{Korreck2005}, however, predict that the net heating due to a fast neutral precursor is too small to account for the observed narrow component line broadening. In contrast to a CR precursor, the efficiency of preshock heating by fast neutrals is sensitive to the degree of thermal equilibration between electrons and ions, the shock speed, and the neutral fraction upstream. It is therefore difficult to explain the narrow range in line broadening seen in many Balmer-dominated filaments in which diverse shock conditions obtain.

The possibility that the narrow component flux increase ahead of the gas subshock is due to a superposition of multiple shocks in the line of sight has not been ruled out. However, as L07 have argued, a superposition of shocks would imply a gradual flux increase of the broad component ahead of the gas subshock in the same manner as the flux of narrow component. This is not observed.

\section{Conclusions}\label{sec:conclusions}
In summary, CR acceleration in the forward shocks of SNRs results in the heating and acceleration of the preshock medium which may explain some features of the optical emission of Balmer dominated filaments. We have found a transient state in the evolution of a shock from one that is initially not modified by CRs to one that is CR dominated, for which the calculated \Ha{} emissivity profile matches the emissivity profile across the Balmer-dominated filament in knot g observed by \citet{Lee-etal2007}. The values of the parameters for this shock model are an initial shock speed $\vs=2000\kms$, a distant upstream CR pressure to thermal gas pressure ratio $\phi_0=1$, a diffusion coefficient $\kappa=2\times10^{24}\cms$, an energy transfer time-scale due to the acoustic instability $\tau=426\yr$, and a lower limit to the injection parameter $\epsilon=4.2\times10^{-3}$. 

The structure of steady shocks that belong to the low CR acceleration efficiency branch of solutions for fast shocks are relatively insensitive to the values of $\tau$, $\epsilon$, $\kappa$, and $\vs$ in the range $300\kms<\vs<3000\kms$, provided that the parameters are chosen such that a steady solution in the low CR acceleration efficiency branch exists. The solutions are usually reached as time-asymptotic states within less than $100\yr$, even if $\epsilon=0$. The mild heating of the preshock gas up to typically $2$--$6\times10^4\Kv$, and the negligible bulk acceleration of the flow in the precursor $(\Delta{}u\leq10\kms)$ may provide a natural explanation for the characteristic broadening of the narrow component linewidth that is observed to lie in the small range $\mathrm{FWHM}=30$--$50\kms$ in many SNRs, and the lack of bulk Doppler shift of the narrow component observed for these cases.

\acknowledgements
The authors would like to thank the referee for carefully reviewing the manuscript and providing feedback that has lead to a substantial improvement of the paper. AYW is grateful for financial support from the Smithsonian Institution Scholarly Studies Fund, and the hospitality during the course of this work at the CfA. This work was funded by HST grant number GO-10577 to the Smithsonian Institution.

\appendix
In the following we give an example of the limits for $\tau$ and $\epsilon$ within which the low acceleration efficiency branch of steady CR modified shocks, as described in Sect.~\ref{sec:steady}, exist. 

If $\epsilon=0$, the exact value for $\phi_c$ depends only on the dimensionless quantity $\eta=\tau\vs^2/\kappa$. The method to find the range in $\eta$ for which two distinct solutions exist for a given distant upstream state is described in \citet{WFH2007} for radiative shocks of lower Mach number. Here, we find, for example, that for shocks for which $\vs=2000\kms$, $\phi_0=1$, $\kappa=2\times10^{24}\cms$, and $\epsilon=0$, the low CR acceleration efficiency branch of solutions exists up to $\tau_a\approx634\yr$, and the (smooth) high CR acceleration efficiency branch of solutions exists down to $\tau_b\approx63\yr$. In the range $\tau_b<\tau<\tau_a$ both an inefficient and an efficient steady solution are allowed. 

As expected, a large injection parameter drives the solution towards one that is CR dominated. For example, if $\vs=2000\kms$, $\kappa=2\times10^{24}\cms$, $\phi_0=1$, and $\tau=426\yr$, the critical value for $\epsilon$ above which a low CR acceleration efficiency solution cannot exist is $\sim1.0\times10^{-4}$. Conversely, a shock for which $\vs$, $\kappa$, $\phi_0$ are the same as above but $\epsilon=4.2\times10^{-3}$ (the value obtained from the model in Sect.~\ref{sec:trans}) can only remain in the inefficient branch of solutions if $\tau\lesssim63$.

\bibliographystyle{hapj}
\bibliography{bib_phd}

\end{document}